 \documentclass{IEEEtran}
\ifCLASSINFOpdf
\else
\fi
\usepackage{multirow}
\usepackage{amsbsy,amsfonts,amsfonts,amssymb,amsmath,epsfig}
\usepackage{graphicx}
\usepackage{graphics}
\usepackage[dvipsnames]{xcolor}
\usepackage{epstopdf}
\usepackage{array}
\usepackage{cite}

\usepackage{array}
\newcolumntype{P}[1]{>{\centering\arraybackslash}p{#1}}

 \usepackage{algorithmic}
  \usepackage{algorithm}

\usepackage[cmintegrals]{newtxmath}
\ifCLASSOPTIONcompsoc
\else
\fi
\begin{document}
%
\title{Deep-Learning Assisted IoT Based RIS for Cooperative Communications}
%
%
%

\author{Bulent Sagir,~ Erdogan Aydin and Haci Ilhan,  \IEEEmembership{Senior Member,~IEEE}
\thanks{B. Sagir is with Turk Telekomunikasyon A. S, 34660, Istanbul, Turkey e-mail: (bulent.sagir@turktelekom.com.tr).}
\thanks{E. Aydin is with the Department of Electrical and Electronics
	Engineering, Istanbul Medeniyet University, Istanbul 34857, Turkey (e-mail:
	erdogan.aydin@medeniyet.edu.tr) (Corresponding author: Erdogan Aydin.)}
	\thanks{H. Ilhan is with Y{\i}ld{\i}z Technical University, Department of Electronics and Communications Engineering, 34220, Davutpasa, Istanbul, Turkey (e-mail: ilhanh@yildiz.edu.tr).}
 }

\maketitle

\begin{abstract}
Reconfigurable intelligent surfaces (RISs) are software-controlled passive devices that can be used as relay ($R$) systems to reflect incoming signals from a source ($S$) to a destination ($D$) in a cooperative manner with optimum signal strength to improve the performance of wireless communication networks. The configurability and flexibility of an RIS deployed in an Internet-of-Things (IoT)-based network can enable network designers to devise stand-alone or cooperative configurations that have considerable advantages over conventional networks. In this paper, two new deep neural network (DNN)-assisted cooperative RIS models, namely, DNN$_R$-CRIS and DNN$_{R, D}$-CRIS, are proposed for cooperative communications. In DNN$_R$-CRIS model, the potential of RIS deployment as an IoT-based relay element in a next-generation cooperative network is investigated using deep learning (DL) techniques for RIS phase optimization. In addition, to reduce the maximum likelihood (ML) complexity at $D$, a new DNN-based symbol detection method is presented with the DNN$_{R, D}$-CRIS model combined with DNN-assisted phase optimization. For a different number of relays and receiver configurations, the bit error rate (BER) performance results of the proposed DNN$_R$-CRIS and DNN$_{R, D}$-CRIS models and traditional cooperative RIS (CRIS) scheme (without a DNN) are presented for a multi-relay cooperative communication scenario with path loss effects. It is revealed that the proposed DNN-based models show promising results in terms of BER, even in high-noise environments with low system complexity.
\end{abstract}


%
\IEEEpeerreviewmaketitle

\vspace{-0.3cm}

\section{Introduction}
%
%
%
%
\IEEEPARstart{F}{orthcoming} 5G and 
more advanced
 wireless network architectures promise a several fold increase in transmission capacity and much lower latency levels compared to current wireless networks by introducing new radio spectra, including millimeter waves, and a bundle of emerging technologies, such as massive multiple-input multiple-output (MIMO), small cells, edge computing, device centric architectures and new channel coding techniques \cite{Boccardi1}, \cite{Coll1}.

The expected improvements in the physical layer apart from the new spectrum usage depend largely on the control of the propagation environment with techniques, such as employing antenna diversity in transmitters and receivers, cooperative relaying schemes and intelligent surfaces. Among these, reconfigurable intelligent surfaces (RIS), which may also be referred to as software-defined surfaces (SDS) or large intelligent surfaces (LIS) in the literature, is a brand new concept that enables every surface in a propagation environment to be used effectively to control the environment for efficient transmission and even to exploit the scattering waves \cite{Basar1,Basar2}. An RIS is a thin surface of multiple reflecting elements made of low-cost passive electronic devices to alter the incident wave phase angles. This concept was first considered in \cite{Subrt1} in 2012, where intelligent walls with active frequency selective surfaces were proposed for controlling the propagation environment in indoor scenarios. In \cite{Hu1}, LIS was proposed to form a contiguous surface of an electromagnetically active material as an alternative to massive MIMO.

Although intelligent surfaces were initially mainly intended for indoor environments, in \cite{Subrt1,Hu1,Huang2,Wu1}, these surfaces were used in outdoor scenarios in the form of passive intelligent mirrors (PIMs) for multiuser multiple-input single-output (MISO) downlink communication \cite{Huang1}; PIMs could be deployed around building exteriors to replace base stations (BS) serving as mobile terminals. The outdoor usage scenarios can be diversified with the progress in material technology.

As an RIS comprises low-cost passive reflectors, one of its significant benefits is that it can be used in a relatively high-cost next-generation network. Its ability to be applied on various surfaces allows it to be used in diverse cases. In \cite{Basar3}, an RF transmitter was deployed in tandem with an RIS for modulating an RF signal, and the RIS was effectively transformed into an RF modulator. Furthermore, various applications of RISs, such as using RISs to create a virtual line-of-sight (LOS) link between BSs and users where it is not feasible to establish a direct reliable link, improving the physical layer security in a wireless network, or realizing simultaneous wireless information and power transfer (SWIPT) in an internet-of-things (IoT) network \cite{Wu1}, \cite{Wu2}, have been investigated. Additionally, RIS-based space shift keying (SSK), spatial modulation (SM) and media-based modulation (MBM) schemes were proposed in \cite{SALAN2021153713, Onal2021}.

Another noteworthy approach is the concept of software-controlled hypersurfaces, which interact with electromagnetic waves in a controlled manner to re-engineer the impinging waves to obtain the desired response \cite{Liaskos8466374}, \cite{Liaskos_2_8449754}. In the proposed model, each hypersurface is made of intelligent surfaces consisting of ultrathin meta atoms that receive commands from a centralized controller and adjust their EM behavior through an 
Internet-of-Things (IoT) 
gateway. The software-configurable nature of hypersurfaces makes it possible to deploy each unit as an IoT device in a broader network structure.

For current and future wireless network architectures, using relay-based cooperative communication techniques to create a link or route diversity is an emerging research area for achieving the desired coverage and throughput capacities \cite{Aydin1,Aydin2,cogen1}. In a typical cooperative system, single or multiple relays are deployed between source ($S$) and destination ($D$) nodes to create alternative routes for signal transmission \cite{Renzo1}. Consequently, for these types of networks, the possibility of using an RIS as a relay ($R$) should be considered, as both devices have very similar functions, e.g., to create route diversity, and turn non-LOS links into LOS links. This is one of the active research areas where the performance and cost of both methods have been investigated and compared. Although relaying is a well-established method to achieve this, the requirements can become too steep to be applied in a field where each relay requires a dedicated power supply and front-end circuitry, which consequently increases the total power requirements, costs, and implementation complexity. One of the main issues discussed extensively in \cite{Renzo1}
 was addressed in 
\cite{Bjornson1} for wireless networks operating in high-frequency bands. The power consumption and energy efficiency of decode-and-forward (DF) and amplify-and-forward (AF) relays against RISs were compared, and it was shown that sufficiently large RISs could outperform relay supported systems in terms of data rates, with reduced implementation complexities \cite{Huang33}.

Using deep learning (DL) or machine learning methods to optimize RIS utilization in wireless networks is a brand new approach and has been addressed in several studies, some of which focus on subjects such as symbol estimation \cite{Khan2020DeepLearningaidedDF}, cooperative communication \cite{Akdemir1}, indoor signal focusing \cite{huang2019indoor} or securing wireless communication \cite{ying2020relay}, \cite{Yang2021}. The results achieved with deep learning techniques often show better or at least comparable performance against conventional methods while significantly reducing the computational complexity.

In RIS-assisted communication systems, channel state information (CSI) estimation can be a challenging process, and a great deal of the given studies so far considers the CSI is perfectly known at the RIS or receiver. However, CSI estimation is possible with techniques already covered in various works \cite{ChannelEstimation1_wang}, \cite{ChannelEstimation1_wei}  beside some other studies dealing with totally blind or imperfect CSI conditions \cite{Basar3}, \cite{Khan2020DeepLearningaidedDF}, \cite{EfficientMIMOdetection_chen}. Deploying DL-based estimation techniques is shown to provide promising results, which can be a groundwork for future research.

\vspace{-0.3cm}
\subsection{Contributions}
In this work, we investigate the possibility of a novel usage scenario: using an RIS as the sole relaying element in a cooperative network, where multiple DL-optimized RISs are utilized in a cooperative configuration between $S$ and $D$. Our work focuses exclusively on the outdoor scenario, where building facades are covered with RISs in a dense urban environment with slow fading channels. In our proposed deep learning-assisted cooperative RIS models (DNN$_R$\:-\:CRIS and DNN$_{R, D}$\:-\:CRIS), RISs are optimized through a deep neural network (DNN), which is trained with the channel state information (CSI) obtained for incident and reflected signals to estimate the phase adjustments required for optimal signal transmission of RISs. Furthermore, in the DNN$_{R, D}$\:-\:CRIS model, instead of using a conventional maximum likelihood (ML) detector at $D$, we deploy another DNN to estimate the received symbols. Using multiple DNNs on different parts of the network, it finally becomes possible to obtain an approximation of an end-to-end DNN-reinforced communication network and to compare its performance against conventional network architectures.

In the proposed DNN optimized model, all DNNs deployed on the RISs are considered to be embedded in the RIS controller hardware and interconnected through an IoT network to communicate and coordinate with other network components. Here, IoT integration is supposed to provide a control platform for all RIS-based relays to supply the system with network and problem management capabilities along with other functional capabilities, such as transmission through relay selection.

The performances of the DNN$_R$\:-\:CRIS/DNN$_{R, D}$\:-\:CRIS models are analyzed in terms of the bit error rate (BER) for an $M$-array quadrature amplitude modulation (QAM) scheme using various relay configurations with path loss effects. Furthermore, a complexity analysis is also included, comparing DNN-assisted and conventional methods. Simulating and measuring the effects of multiple RISs on the received signal in a dense urban environment are the main goals of our work; consequently, the potential of DNN-assisted RIS usage in a next-generation wireless network is exhibited.

For all the scenarios and configurations covered in this work, it is assumed that the channel state information (CSI) is perfectly known at the IoT-based relays interconnected through the IoT network, gathering full CSI data extracted from the pilot signals.

As a result of the analyses, we can summarize the contributions of our paper as follows.
\begin{enumerate}
    \item In the proposed DNN$_R$\:-\:CRIS and DNN$_{R, D}$\:-\:CRIS models, RISs deployed as IoT-based relay elements in a cooperative network are presented. For the RIS-based relaying model, it is possible to select the best performing $R$ among multiple relay configurations between $S$ and $D$ with a choice of conventional relay selection methods.
    \item RIS-based relays are conceived as IoT devices communicating in a broader IoT network where DNN input signals and other vital signaling are transmitted through IoT gateways. This kind of IoT integration for RIS-based relays in a cooperative communications scenario is also limited in the literature.
    \item In both models, RIS configurations are optimized with DNNs. A DNN is deployed for controlling each RIS in the system and to optimize the phase adjustments in real time. In the DNN$_{R, D}$\:-\:CRIS model, an additional DNN is deployed at the destination for symbol detection instead of an ML detector. The performance results of the ML detector and DNN at the destination are compared in terms of BER. Additionally, through this model, DNN-assisted phase and symbol detection are collectively deployed and analyzed in a cooperative network.
    \item In an RIS-based multiple relay case, the effect of the relay locations on receiver performance with path loss effects is also investigated. The distance from the relays to $S$ or $D$ in conjunction with fading channels has a critical impact on BER performance, and this is demonstrated extensively in our work.
    \item An extensive evaluation of the results is also performed with a complexity analysis, comparing proposed DNN-assisted and conventional methods.
\end{enumerate}

	\vspace{-0.4cm}

\subsection{Organization and Notations}
\begin{figure}[t!]
\centering
\includegraphics[width=0.48\textwidth]{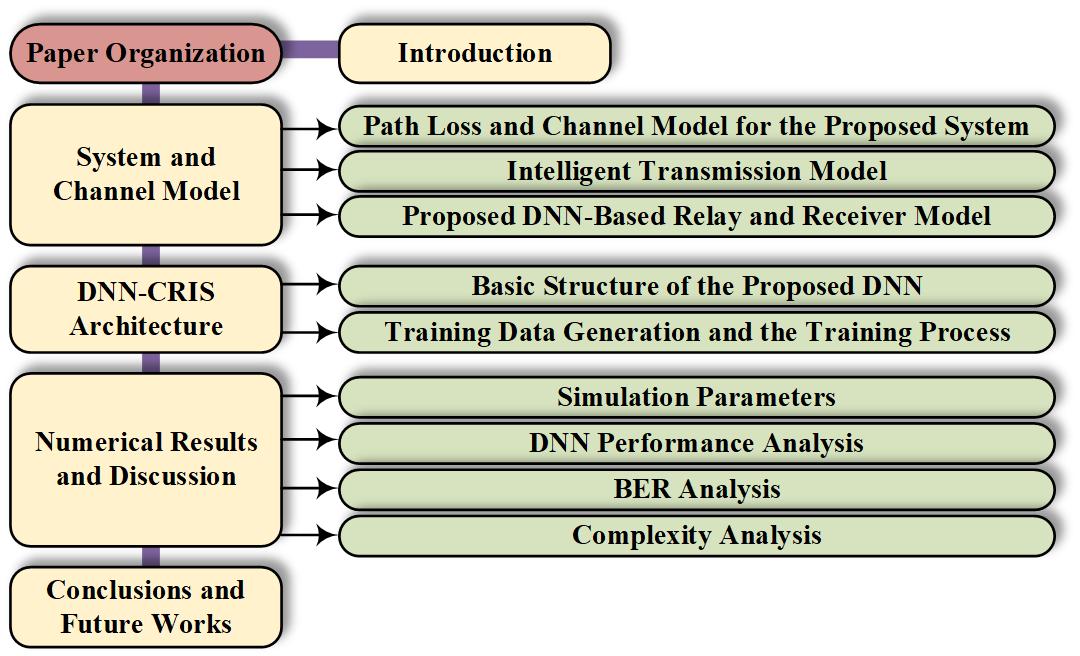}
\caption{Organization of the paper.}
\label{org_1}
	\vspace{-0.3cm}
\end{figure}

The organization of this paper is given in Fig. \ref{org_1}. The system and channel models are presented in Section II, with the theoretical background of transmission and the proposed DNN-based relay and receiver models. In Section III, the architecture of the proposed model with details of the basic structure of the proposed DNNs, training data generation and training processes are thoroughly covered for both DNNs. Section IV provides simulation setup and performance results for a series of RIS configurations, including a detailed performance comparison of the proposed model against various parameters. Finally, Section V concludes the article.

\textit{Notations}: The following notations are used throughout this paper. \textit{i}) Bold lower/upper case symbols represent vectors/matrices; \textit{ii}) $(\cdotp)^T$ and $\left\|\cdotp \right\|_F $  denote transpose and Frobenius norm operators, respectively; \textit{iii}) $\Re(.)$ and $\Im(.)$ are the real and imaginary parts of a complex-valued quantity; \textit{iv}) $(.)^*$ represents the conjugate operator. $\textit{v}) [.]_{a\times b}$ indicates the dimensions of a matrix with $a$ rows and $b$ columns; \textit{vi}) $\mathcal{O}(.)$ represents the time or computational complexity of an algorithm, in terms of number of operations.

\begin{figure}[t!]
	\centering
	\includegraphics[width=0.48\textwidth]{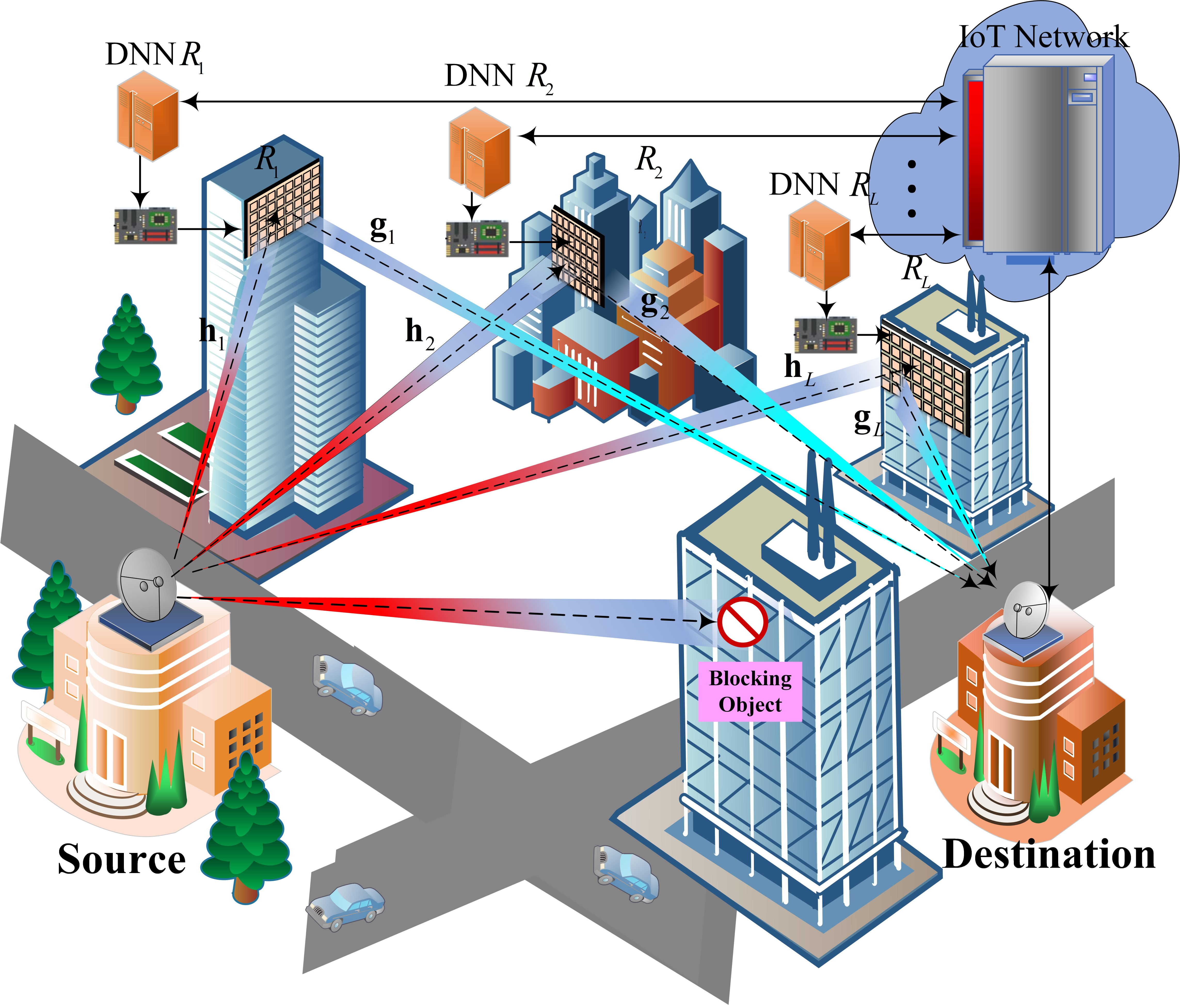}
	\caption{System model of the DNN$_R$\:-\:CRIS/DNN$_{R, D}$\:-\:CRIS.}
	\label{system_model}
 	\vspace{-0.4cm}
\end{figure}

\vspace{-0.3cm}

\section{Transmission Through DNN$_R$\:-\:CRIS/DNN$_{R, D}$\:-\:CRIS:  System Model }

In this section, we provide an overview of the generic model of the proposed DNN$_R$\:-\:CRIS and DNN$_{R, D}$\:-\:CRIS schemes. The system model under consideration will be named as DNN$_R$\:-\:CRIS if a DNN is used in $R$ only, and DNN$_{R, D}$\:-\:CRIS if DNN is used in both $R$ and $D$.

System model for the proposed scheme is shown in Fig.  \ref{system_model}, where $S$ and $D$ contain a single transmit and receive antennas. In the considered scenario, assuming no LOS link between $S$ and $D$, cooperative transmission occurs only through multiple RIS elements. We assume that each RIS has the form of a reflect-array comprising $N$ simple and reconfigurable reflector elements and acts as a $R_\ell$ in a cooperative network routing the transmission from $S$ to $D$. In the proposed system model, all RISs are assumed to be individually controlled and configured by an IoT based controller, which hosts both management and DNN software for the RIS. Software based DNN constantly computes the phase shifts for each reflector and adjusts the configurations for the reflector elements accordingly. Through the IoT network, each controller can be conceived as a gateway for forwarding input signals to the DNN and managing the RIS based relay for additional functionality, such as activation/deactivation, fault management, troubleshooting, software updates, etc. In the proposed model, each RIS is considered as an IoT device where the existing IoT platforms and communication protocols can be deployed, details of which will be beyond the context of this work. The software based DNN architecture will be covered in detail in subsequent sections. 

At the destination end, two scenarios are considered. First, an ML-detector is deployed to estimate the received symbols (DNN$_R$\:-\:CRIS). Second, another DNN is used to estimate the received symbols implemented in the receiver hardware, effectively replacing an ML-detector (DNN$_{R, D}$\:-\:CRIS). We'll evaluate and compare the performances of both detection schemes in numerical results section.

\begin{figure}[t!]
	\centering
	\includegraphics[width=0.48\textwidth]{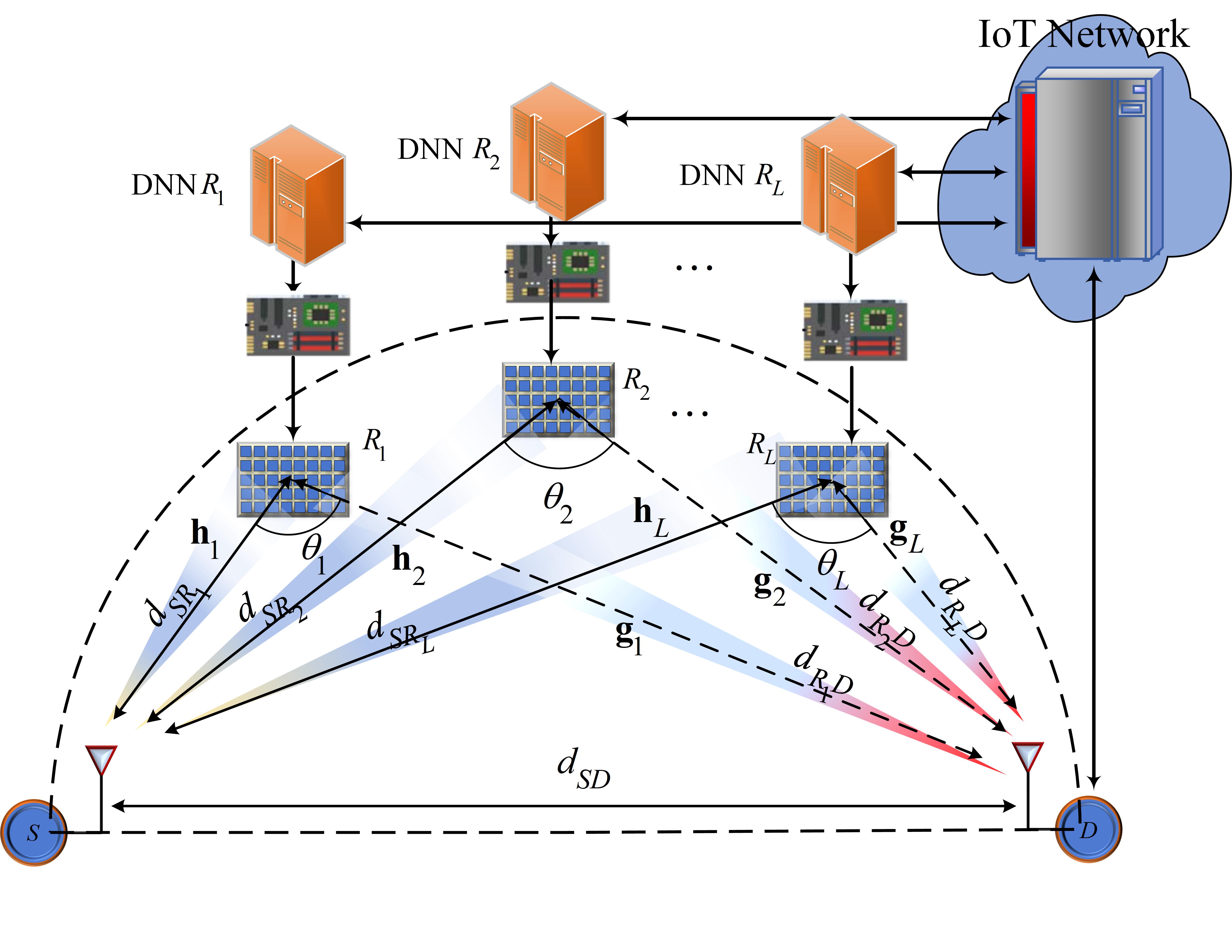}
 \vspace{-0.3cm}
	\caption{Channel model for DNN-CRIS system.}
	\label{S_model1}
 \vspace{-0.3cm}
\end{figure}

\vspace{-0.3cm}
\subsection{Path Loss and Channel Model for the Proposed System}

For the transmission model, we assume that $S$, $D$ and each $R_\ell$ are positioned in a triangle formation in a two dimensional space as shown in Fig. \ref{S_model1} where $d_{SD}$, $d_{SR_\ell}$ and $d_{R_{\ell}D}$ represent the distances of source to destination ($S \to D$), source to $\ell^{th}$ relay ($S \to R_\ell$) and $\ell^{th}$ relay to destination ($R_\ell \to D$), respectively, where $\ell=1,2,\ldots,L$ and here $L$ is the number of relays. In this configuration, $S$ and $D$ stay at the either end of the triangle base and $R_\ell$ is positioned on the top corner such that all three corners are kept within the half-circle as shown in Fig. \ref{S_model1}, assuming  $ \pi/2<\theta_\ell<\pi$ where $\theta_\ell$ is the angle between $d_{SR_\ell}$ and $d_{R_{\ell} D}$. In this formation, $R_\ell$ lies on the circle arc when $\theta_\ell=\pi/2$, and lies on the line $S \to D$ when $\theta=\pi$. In all cases, $d_{SD}$ will be equal to the diameter.

In the proposed channel model, all links are assumed to be exposed to both long-term free space path loss and short-term Rayleigh fading. Here, the path loss is proportional to $d^{-c}$ where $d$ is the propagation distance and $c$ is the path loss exponent \cite{Rappaport1992}. As the path loss for the channel $S \to D$ is assumed to be unity, the relative gains of $S \to R_\ell$ and $R_\ell \to  D$ can be defined as $G_{SR_\ell}=(d_{SD}/d_{SR_\ell})^c$ and $G_{R_{\ell}D}=(d_{SD}/d_{R_{\ell}D})^c$ respectively. Here, $G_{SR_{\ell}}$ and $G_{R_{\ell}D}$ are related by law of cosines, which is \cite{Proakis2007}
\begin{eqnarray}\label{eq_cosines}
G_{SR_{\ell}}^{-2/c}+G_{R_{\ell}D}^{-2/c}-2G_{SR_{\ell}}^{-1/c}G_{R_{\ell}D}^{-1/c}\cos\theta_{\ell}=1.
\end{eqnarray}
Through (\ref{eq_cosines}), assuming $d_{SR_{\ell}}$ and $\theta_{\ell}$ are given and $d_{SD}=1$, $d_{R_{\ell}D}$ can be computed in order to apply path loss to complex channel coefficients between $S \to R_{\ell}$ and $R_{\ell} \to D$, which can be expressed as  $\textbf{h}_\ell= \big[h_{\ell,1}, h_{\ell,2},\ldots,h_{\ell,N}\big] \in \mathbb{C}^{1 \times N}$ and $\textbf{g}_\ell= \big[g_{\ell,1}, g_{\ell,2},\ldots,$ $g_{\ell,N}\big] \in \mathbb{C}^{1 \times N}$, where $\ell=1,2,\ldots,L$ and $\mathbb{C}$ denotes the set of complex numbers. Channel vector elements  can be expressed in terms of channel amplitudes and phases $ h_{i,j} = \alpha_{i,j}e^{-j\varphi_{i,j}}$ and  $ g_{i,j} = \beta_{i,j}e^{-j\theta_{i,j}}$, respectively where $i\in\big\{ 1,2,\ldots,L \big\}$ and $j\in\big\{ 1,2,\ldots,N \big\}$. In the considered system structure, all channel coefficients are assumed to be Rayleigh fading distribution. Under this assumption, we have $h_{i,j},  g_{i,j} \sim \mathcal{CN}(0,1)$, where $h_{i,j},  g_{i,j} \sim \mathcal{CN}(0,\sigma^2)$ stands for the complex Gaussian distribution with zero mean and $\sigma^2$ variance.

With this model, it's possible to observe the effect of various relay positionings and evaluate their performances.

\subsection{Intelligent Transmission Model}

The noisy received baseband  signals  reflected through the $\ell^{th}$ relay with $N$ passive elements   can be expressed at the $D$ as follow:
\begin{eqnarray}\label{eq_y_l} 
y_\ell &=& \sqrt{G_{SR_{\ell}}G_{R_{\ell}D}}\Bigg( \sum_{n=1}^{N}h_{\ell,n}e^{j \phi_{\ell,n}}g_{\ell,n} \Bigg)x + w_\ell  \nonumber \\
&=& \sqrt{G_{SR_{\ell}}G_{R_{\ell}D}} \bigg(\textbf{h}_\ell\boldsymbol{\Phi}_\ell\textbf{g}_\ell^T\bigg)x + w_\ell,
\end{eqnarray}
Here,  $x$ is the modulated $M$-QAM signal, and $w_\ell \sim \mathcal{CN}(0,\mathcal{N}_0 )$ represents the additive white Gaussian noise (AGWN) with with zero mean and $\mathcal{N}_0/2$ variance per dimension at the receiver \cite{Coop_comms_Liu}. $\boldsymbol{\Phi}_\ell \triangleq \mathtt{diag}\big(\boldsymbol{\phi}_\ell\big) \in \mathbb{C}^{N \times N}$ is the diagonal phase matrix for each RIS including the adjusted phase angles by RIS reflecting elements,
where the phase vector for $\ell^{th}$ RIS is defined as $\boldsymbol{\phi}_\ell=\Big[\phi_{\ell,1},\phi_{\ell,2},\dots,\phi_{\ell,N}\Big]$. The vectorial representation of the noisy baseband signals reflected through all relays with $N$ passive elements can be expressed at $D$ as follows:
\begin{eqnarray}\label{eq_y_vec} 
\textbf{y} &=& \Big[y_1, y_2,\ldots,y_L\Big]^T \in \mathbb{C}^{L \times 1}. 
\end{eqnarray}
From (\ref{eq_y_l}), we can find the instantaneous signal-to-noise ratio (SNR) at $D$ for the reflected signals through the $\ell^{th}$ relay with $N$ passive elements as
\begin{equation}\label{eq_SNR}
\gamma_\ell  = \frac{{{{\left| {\mathop \sum \nolimits_{i = 1}^N {\alpha_{\ell,i}}{\beta_{\ell,i}}{e^{j\left( {{\phi_{\ell,i}} - {\varphi_{\ell,i}} - {\theta_{\ell,i}}} \right)}}} \right|}^2}{G_{SR_{\ell}}G_{R_{\ell}D}E_s}}}{{{\mathcal{N}_0}}},
\end{equation}
where $E_s$ is the average transmitted energy per symbol. Adjustment of the phase angles by RIS is based on the fact that the maximum signal strength at $D$ can be achieved when \({\phi _{\ell,i}} = {\varphi _{\ell,i}} + {\theta _{\ell,i}}\) for \(i = 1, \ldots ,N\), assuming the channel phases are known at the RIS. This can be verified by the identity \({\Big| {\mathop \sum_{i = 1}^N {v_i}{e^{j{\xi _i}}}} \Big|^2} = \mathop \sum_{i = 1}^N v_i^2 + 2\mathop \sum_{i = 1}^N \mathop \sum_{t = i + 1}^N {v_i}{v_t}\cos \left( {{\xi _i} - {\xi _t}} \right)\) where maximum value is achieved when \({\xi _i} = {\xi _t}\) for all $i$. Using this fact for phase adjustment at each reflecting element, the maximum instantaneous SNR is achieved at $D$ for the reflected signals through the $\ell^{th}$ relay with $N$ passive elements as
\begin{equation}\label{eq_SNR_max}
\gamma_\ell  = \frac{{{{\left| {\mathop \sum \nolimits_{i = 1}^N {\alpha_{\ell,i}}{\beta_{\ell,i}}} \right|}^2}{G_{SR_{\ell}}G_{R_{\ell}D}E_s}}}{{{\mathcal{N}_0}}} 
= \frac{{{\chi^2}{G_{SR_{\ell}}G_{R_{\ell}D}E_s}}}{{{\mathcal{N}_0}}},
\end{equation}
where $\chi= \big| {\mathop \sum \nolimits_{i = 1}^N{\alpha_{\ell,i}}{\beta_{\ell,i}}} \big|$.  The total  instantaneous  SNR ($\gamma_{T}$)  at $D$ for the reflected signals through all relays with $N$ passive elements can  be written   as $\gamma_{T}  = \sum_\ell^L\gamma_{\ell}$.

Finally, the achievable throughput performance of the considered
DNN$_R$\:-\:CRIS/DNN$_{R, D}$\:-\:CRIS communication systems is given by $\mathcal{R}:= \log_2(1+\gamma_{T} )$.

\subsubsection{Combining Techniques at the Destination}
Combining techniques used at $D$ are as follows.

\textit{i) Transmitting Through the Best Performing Relay}:
In this technique, the relay that provides the lowest BER value is selected among $L$ relays for transmission. In this case, the best relay selection (RS) criterion will be \cite{Coop_comms_Liu},
\begin{eqnarray}\label{eq_BER_rs}
\widetilde{BER}_{\ell^*}=\arg\min_\ell\{BER_\ell\},\quad \ell=1,2,...,L.
\end{eqnarray}
Here, $\ell^*$ denotes the relay which provides the lowest BER value among  $\ell=1,2,...,L$ relays.

 \begin{figure*}[t!]
	\centering
	\includegraphics[width=0.8\textwidth,height=0.4\textheight]{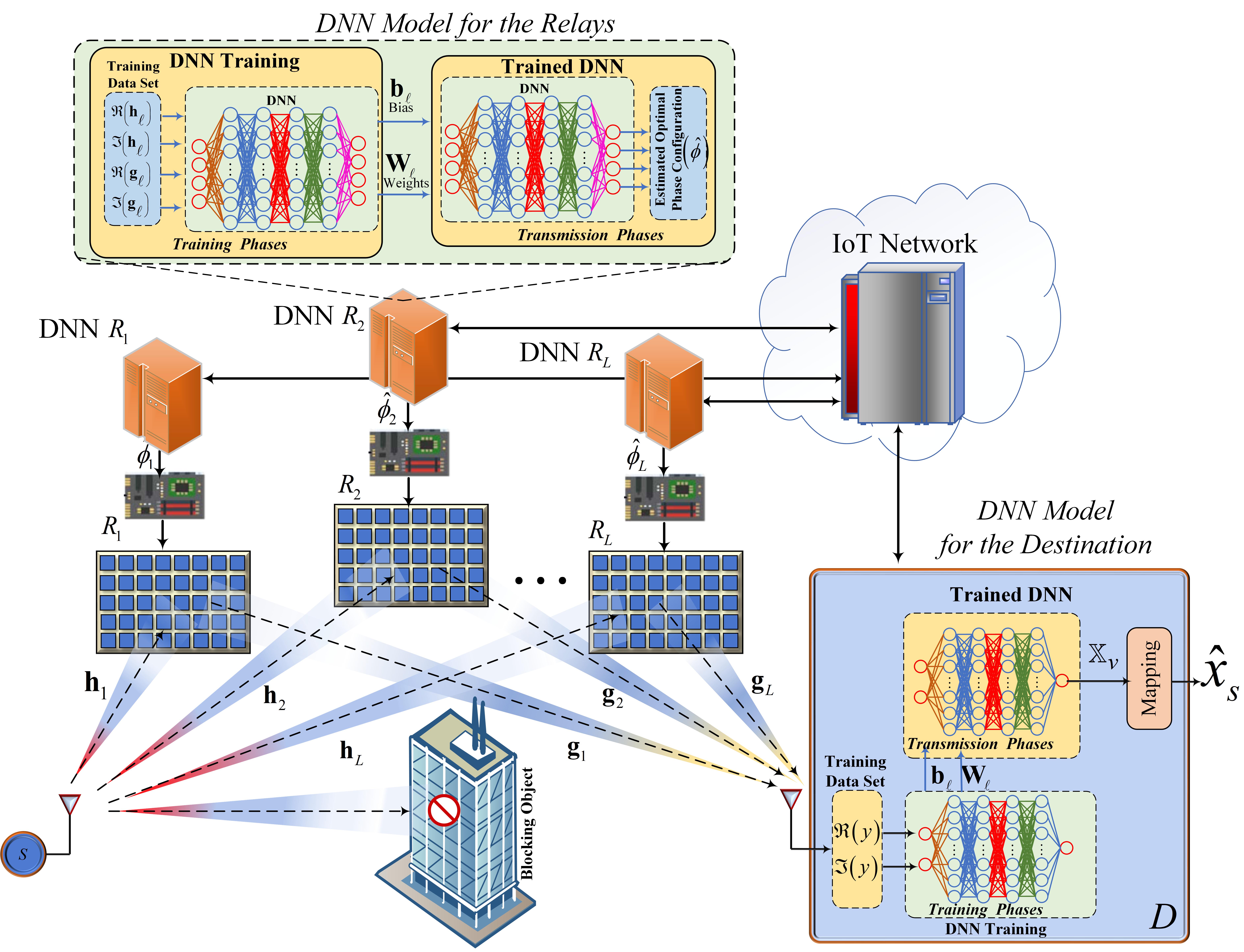}
	\caption{The training and transmission phases of the proposed deep learning assisted CRIS scheme.}
	\label{DNN_model1}
 \vspace{-0.3cm}
\end{figure*}

\textit{ii) Maximum Ratio Combining (MRC)}:
Using MRC technique for combining the signals received through multiple relays, the combined overall signal at $D$ can be written as,
\begin{eqnarray}\label{eq_y_MRC}
y_{MRC}=\zeta_1 y_1 + \zeta_2 y_2+...+\zeta_L y_L,
\end{eqnarray}
where $y_\ell$ is defined in (\ref{eq_y_l}). 
Since the product of a complex number with its conjugate cancels the phase and results with the sum of the squared value of the magnitudes, the combining coefficient $\zeta_\ell$ should be conjugate of the complex channel response so that the product will yield absolute squares of the channel responses. Thus, the combining coefficient $\zeta_\ell$ for the $\ell^{th}$ relay will be expressed as follows \cite{Coop_comms_Liu}:
\begin{eqnarray}\label{eq_w_l}
\zeta_\ell &=& \Bigg(\sum_{n=1}^Nh_{\ell,n}e^{j \phi_{\ell,n}}g_{\ell,n}\Bigg)^* = \ (\mathbf{h}_\ell\boldsymbol{\Phi}_\ell\mathbf{g}^T_\ell)^*. 
\end{eqnarray}
After MRC based combining is carried out at the receiver, ML detection applied on the combined signal can be expressed as,
\begin{eqnarray}\label{eq_x_MRC} 
	\hat{x}_{MRC} &=& \mathrm{arg}\ \underset{\upsilon \in \{1,2,\ldots,M\}}{\min}\!\Biggl\{\Bigg|y_{MRC}\!-\!\Bigg(\!\zeta_1 \! \! \sum_{n=1}^{N}h_{1,n}e^{\hat{\phi}_{\ell,n}}g_{1,n} \nonumber \\
	\! \!\! \!\! \!\! \!\! \!\! \!\! \!\! \!\! \!\! \!\! \!\! \!\! \!\! \!\! \!\! \!\! \!\! \!\! \!\! \!\! \!&  &\! \!\! \!\! \!\! \!\! \!\! \! \!\! \!\! \!\! \!\! \!\! \!\! \!\! \!\! \! \!\! \!\!\! \!+\zeta_2\sum_{n=1}^{N}h_{2,n}e^{\hat{\phi}_{\ell,n}}g_{2,n}+...+\zeta_L \sum_{n=1}^{N}h_{L,n}e^{\hat{\phi}_{\ell,n}}g_{L,n}\Bigg)x_\upsilon\Bigg|^2\Biggr\}.
\end{eqnarray}
Here, $M$ is the modulation order and $x_v$ is the symbol to be transmitted where $v\in{1,2,...,M}$.

\textit{iii) Maximum Likelihood (ML) Detector}: For the first scenario in the proposed system model, an ML detector is deployed at $D$ for symbol detection. Using ML detector, the estimated symbol at the receiver can be defined as \cite{Proakis2007},
\begin{eqnarray}\label{eq_x_ML} 
	\hat{x}_{ML} &=& \mathrm{arg}\ \underset{\upsilon \in \{1,2,\ldots,M\}}{\min} \Biggl\{\Bigg{\lVert}\mathbf{y}-\Bigg[ \sum_{n=1}^{N}h_{1,n}e^{\hat{\phi}_{\ell,n}}g_{1,n}\nonumber \\
	& & \!\!\!\!\!\!\!\!\!\!\!\!\!\!\!\!\!\!\!\!\!\!\!\sum_{n=1}^{N}h_{2,n}e^{\hat{\phi}_{\ell,n}}g_{2,n}... \sum_{n=1}^{N}h_{L,n}e^{\hat{\phi}_{L,n}}g_{L,n}\Bigg]^Tx_\upsilon\Bigg{\rVert}^2\Biggr\}.
\end{eqnarray}

\subsection{Proposed DNN-Based Relay and Receiver Model}

This subsection explains how DNN assisted RIS technique can be implemented in a RIS-assisted cooperative communication scenario.

Artificial neural networks (ANNs) are mathematical structures inspired by the biological neural networks consisting of interconnected neurons to form a network that can be trained to perform desired tasks. A DNN can be realized with an ANN with multiple hidden layers with a much greater function approximation ability than a single hidden layer network \cite{Deep_Learning_Lecun}. In the proposed model, deep neural networks have been deployed at two critical locations: At the relays to adjust the phase shifts for each reflector and at $D$ to estimate the received symbols.

As stated earlier, each DNN based relay or RIS in the system is considered to be equipped with an IoT based controller unit responsible for controlling the states of reflecting elements on the RIS to adjust the reflection phase shifts for outgoing signals. In this respect, proposed DNN software will be implemented in the controller hardware. The controller is also responsible for communicating with the outer world, performing as a gateway for both network management functions and last but not least, for receiving CSI data to be fed into the DNN, critical for both training and transmission phases, which will be discussed in the following section.

\begin{figure*}[t!]
	\centering
	\includegraphics[width=0.85\textwidth]{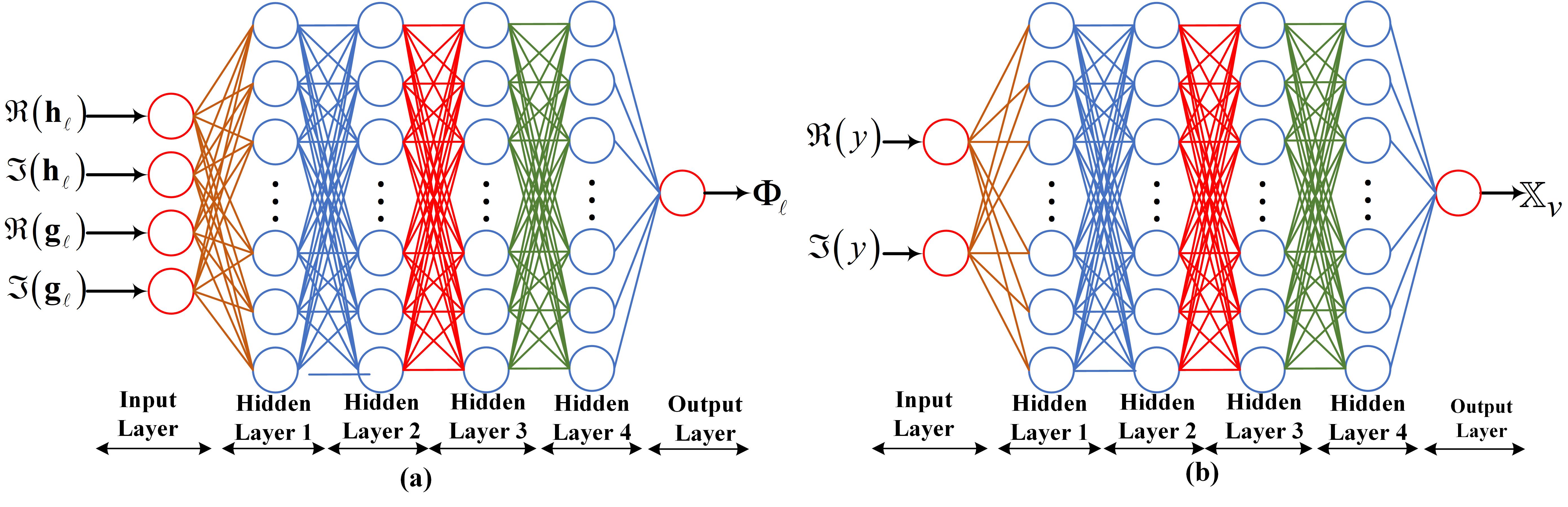}
    \caption{Proposed DNN architectures (a) at the relay and (b) at the destination, for DNN$_R$\:-\:CRIS/DNN$_{R, D}$\:-\:CRIS scheme. }
    \label{DNN architecture}
    \vspace{-0.3cm}
\end{figure*}

 \subsubsection{DNN$_R$\:-\:CRIS Model}
 
In this model, a DNN is deployed in each $R$, pre-trained with a sample set of channel coefficients and respective phase adjustment information $\big(\textbf{h}_\ell,\textbf{g}_\ell,\boldsymbol{\phi}_\ell\big)$ based on the theoretical framework provided in the next section. The objective of the proposed DNN based system is to compute the optimal phase vector that maximizes reflected signals from the relays. The trained DNN dynamically updates the phase adjustments on each RIS reflecting element according to channel information varying in time. Hence, the whole process is divided into two phases: 
\begin{itemize}
	\item [i)]\emph{Training Phase:} As shown in Fig. \ref{DNN_model1}, the training data set is created during training phase, containing $S \to R_\ell$ and $R_\ell \to D$ channel coefficients and the corresponding optimum phase configurations $\big($i.e.,  $\textbf{h}_\ell,\textbf{g}_\ell,\boldsymbol{\phi}_\ell$, where $\ell=1,2,\ldots,L \big)$. Here, the CSI is assumed to be perfectly obtained by the pilot signals through IoT gateways, and pre-processed by the controller hardware to compile the training data set. After the data set is created, the elements of this data set will be used to train the DNNs to find optimum RIS phase vectors for each $R$, and consequently, weight and bias coefficients ($\textbf{W}_k,\textbf{b}_k$) of DNN will be updated. 
	\item [ii)]\emph{Transmission Phase:} During actual transmission, DNN receives the new $S \to R_\ell$ and $R_\ell \to D$ channel coefficients and computes the adjusted phase vector and forwards it to the RIS. The RIS adopts the computed phase configuration to assist in the transmission until new channel coefficients received.
\end{itemize}

 \subsubsection{DNN$_{R, D}$\:-\:CRIS Model}

In the second scenario, another DNN is deployed at $D$ instead of an ML detector to detect the received symbols. Like the preceding DNNs at the relay stage, the process is divided into training and transmission phases where the network is first trained using the appropriate training data set. Then the trained network carries out the symbol detection based on the noisy signals received at $D$ during the transmission phase, respectively.
\begin{itemize}
	\item [i)]\emph{Training Phase:} Training data set contains the noisy and path faded signals received at $D$ as feature vectors and the respective modulated symbols as outputs. The objective of the training process is to update the $\textbf{W}_k$ and $\textbf{b}_k$ parameters of DNN such that for every received signal, DNN converges to the desired symbol.
	\item [ii)]\emph{Transmission Phase:} The DNN at $D$ with updated parameters $\textbf{W}_k$ and $\textbf{b}_k$ estimates the correct symbols from the noisy signals reaching $D$ in real-time during the actual transmission.
\end{itemize}

Here, the detection process at $D$ will be defined as a multiclass classification problem such that the elements of the symbol (constellation) set determined by modulation order are mapped to classes to be estimated by the DNN. The details of the process will be covered in the following sections.

\section{DNN$_R$\:-\:CRIS/DNN$_{R, D}$\:-\:CRIS Architecture}

This section presents the architectural details of the proposed DNN schemes to find the optimal phase adjustments for RIS-based relays and detect symbols at $D$. The proposed scenario is based on a double DNN architecture as shown in Fig. \ref{DNN_model1}, where the first DNN is deployed at $R$ for estimating the optimal phase vectors using $S \to R_\ell$ and $R_\ell \to D$  channel coefficients and the second DNN is deployed at $D$ to detect the received symbols using path faded and noisy signals at $D$. The subsections below explain our analysis of the proposed system's structure, training procedure, and activation function. As the model is based on a double DNN architecture, both the common and specific structural properties of both DNNs will be pointed out in the next section for convenience.

	\vspace{-0.3cm}
\subsection{Basic Structure of the Proposed DNNs}

Let $\Theta \triangleq \bigl\{\boldsymbol{\theta}_1,\boldsymbol{\theta}_2,\ldots,\boldsymbol{\theta}_{\mathcal{K}}\bigr\} $ contains $\mathcal{K}$ set of parameters, where $\mathcal{K}$ is the number of layers. A feedforward DNN structure with $\mathcal{K}$ layers is defined as a mapping $\mathcal{F}\bigl(\textbf{u}_0,\boldsymbol{\theta}\bigr): \mathbb{R}^{N_0 \times 1}\mapsto \mathbb{R}^{N_\mathcal{K} \times 1}$ of the input vector $\textbf{u}_0 \in \mathbb{R}^{N_0 \times 1}$ to an output vector in $\mathbb{R}^{N_\mathcal{K} \times 1}$ with the $\mathcal{K}$ iterative processing steps, and  expressed as follows: 
\begin{eqnarray}\label{eq_u_k}
\textbf{u}_k \triangleq f_k\Bigl( \textbf{u}_{k-1}; \boldsymbol{\theta}_k \Bigr), \ k=1,2,\ldots,\mathcal{K}
\end{eqnarray}
where $ f_k\bigl( \textbf{u}_{k-1}; \boldsymbol{\theta}_k \bigr):\mathbb{R}^{N_0 \times 1}\mapsto \mathbb{R}^{N_\mathcal{K} \times 1}$ expresses the mapping implemented by the $k^{th}$ DNN layer. This mapping depends on both a set of $\boldsymbol{\theta}_k $ parameters and the $\textbf{u}_{k-1}$  output vector  from the previous $(k-1)^{th}$ layer. The mapping function can be a function of random variables. Hence,  $f_k \bigl( .;.\bigr)$ may have a stochastic structure.
Since all neurons in the proposed DNN architectures are fully connected to all following neurons,  mapping function $ f_k\bigl( \textbf{u}_{k-1}; \boldsymbol{\theta}_k \bigr)$ can be written via $\boldsymbol{\theta}_\ell \triangleq  \bigl\{\textbf{W}_k,\textbf{b}_k\bigr\}$ as follows:
\begin{eqnarray}\label{eq_f_k}
	f_k\Bigl( \textbf{u}_{\ell-1}; \boldsymbol{\theta}_k \Bigr) = \varsigma\Big(  \textbf{W}_k \textbf{u}_{k-1} + \textbf{b}_k \Big)
\end{eqnarray}
Here, $\textbf{W}_k \in \mathbb{R}^{N_k \times (N_k-1)}  $ and $\textbf{b}_k \in \mathbb{R}^{N_k \times 1} $ refer to the neurons’ weight and bias vectors, respectively. $\varsigma(.)$ denotes an  activation function. It should be stated that the basic mapping structure and processing steps are valid for both DNNs deployed in the system.

Proposed DNN architectures are shown in Fig. \ref{DNN architecture}, where both DNNs consist of $4$ hidden layers, one input and one output layer, and each hidden layer has $256$ neurons with a fully connected structure. In the proposed models, both rectified linear unit (ReLU) and hyperbolic tangent (tanh) activation functions are considered to be deployed at hidden layers. ReLU and tanh functions can be defined as
\begin{eqnarray}\label{eq_relu}
\varsigma_{ReLU}(z)=max(0,z),
\end{eqnarray}
\begin{eqnarray}\label{eq_tanh}
\varsigma_{tanh}(z)  = \frac{\sinh{z}}{\cosh{z}} = \frac{e^z-e^{-z}}{e^z+e^{-z}},
\end{eqnarray}
for a neuron with input $z$\cite{Khan2020DeepLearningaidedDF}, respectively. For both DNNs, we have applied ReLU and tanh activation functions consecutively in order to evaluate accuracy and convergence speed for the particular tasks in Table \ref{table:1}. It is observed that ReLU converged nearly 5 times faster on average than tanh during the training phase showing the same or better accuracy levels.

\begin{table}
\centering
\caption{ReLU and tanh Training Performance Comparison.}
\begin{tabular}{| c | c | c |}
 \hline
 \hline
 \textbf{} &  \textbf{ReLU} & \textbf{tanh} \\
 \hline\hline
 Mini-batch Loss & 0,0385 & 0,0391 \\
 \hline
 Mini-batch RMSE & 0,028 & 0,028 \\
 \hline
No. of Iterations & 60 & 250 \\
 \hline
Time Elapsed & 5'23'' & 25'35'' \\
 \hline
 \hline
\end{tabular}
\label{table:1}
\vspace{-0.3cm}
\end{table}

As the inputs for both DNNs in the model exhibit sequence data characteristics, both DNN's input layers are configured as the sequence input layer so that channel coefficients and received symbols can be used as input sequences for DNNs deployed at the relay and destination, respectively, albeit evaluated as different data types. On the other hand, the most significant difference between the two DNNs is the output layer implementation, which characterizes the network response behavior. The output layer of the DNN at $R$ is configured as a regression layer, where the output will be complex phase adjustment estimations $\hat{\Phi}_\ell$. In contrast, the output layer of the DNN at $D$ is configured as a classification layer for $M$-QAM modulated symbol estimations $\hat{x}$. In a typical classification network, a classification layer follows a softmax layer with the activation function defined as,
\begin{eqnarray}\label{eq_softmax}
\varsigma_{softmax}(\textbf{z})_i = \frac{e^{z_i}}{\sum^K_{j=1}e^{z_j}}.
\end{eqnarray}
Here, $\textbf{z}$ and $z_i$ denote $K$ vectors from previous layer and the elements of the input vector, respectively, where $i=1,...,K$ and $\textbf{z}=(z_1,...,z_k)\in\mathbb{R}^K$ \cite{bridle-softmax}. The objective of using the softmax function is to convert $K$ vectors at the input to $K$ vectors at the output whose total is equal to 1. In this way, those vectors are normalized and mapped onto a probability distribution, and the vector values with the highest probability are transmitted to the next layer for classification.

Overfitting is a common problem in machine learning as the network performs too well on the training data set but performs poorly when it encounters new data. In the considered system, 10\% of the training data set is split and reserved as validation data to evaluate the model during the training process in terms of generalization performance, with remaining data reserved for training.  

\subsection{Training Data Generation and the Training Process}

For the proposed models, the pre-processing of the CSI data involves processing the CSI data before it is applied to DNN as input, where the data is arranged and manipulated to be deployed as feature vectors and then applied to DNN at $D$. In this section, training data generation and training process are explained for DNN$_R$\:-\:CRIS and DNN$_{RD}$\:-\:CRIS models.

\subsubsection{DNN$_R$\:-\:CRIS Model}
Training of the proposed DNN model requires feature vector samples, consisting of CSI data for incoming and outgoing channels $h_{\ell,n}$ and $g_{\ell,n}$, respectively, and adjusted phase angles $\boldsymbol{\phi}_{\ell,n}$ accordingly. Because of the processing constraints of DNN, complex channel coefficients are separated into real and imaginary parts as using the complex CSI data directly as input is not guaranteed to provide consistent results \cite{Matlab_complex}. Preservation of the complete channel information is crucial, so the imaginary part should be processed along with real part, consequently doubling the number of features per sample.
Thus, the resulting feature vector set becomes as,
\begin{eqnarray}\label{eq_F_train}
\mathbf{F}_{\text{train}}^i  = \Big[ \Re\big(h_{\ell,n} \big) \ \Im\big(h_{\ell,n}\big) \ \Re\big(g_{\ell,n}\big) \ \Im\big(g_{\ell,n}\big)\Big]_{1\times4}.
\end{eqnarray}
Here, $\mathbf{F}_\text{train}\triangleq{\mathbf{F}_\text{train}^1,\mathbf{F}_\text{train}^2,\ldots,\mathbf{F}_\text{train}^s}$ for $s$ samples, $i=1,2,...,s$ and $\mathbf{F}_{\text{train}}^i\in\mathbf{F}_{\text{train}}$. Likewise, for each feature vector, the corresponding phase vector $\boldsymbol{\phi}_{\ell,n}$ is separated into real and imaginary parts, forming the output vector as,
\begin{eqnarray}\label{eq_O_train}
\mathbf{O}_{\text{train}}^i  = \Big[ \Re\big(\boldsymbol{\phi}_{\ell,n}\big) \ \Im\big(\boldsymbol{\phi}_{\ell,n}\big)\Big]_{1\times2}
\end{eqnarray}
where $\mathbf{O}_\text{train}\triangleq{\mathbf{O}_\text{train}^1,\mathbf{O}_\text{train}^2,\ldots,\mathbf{O}_\text{train}^s}$ for $i=1,2,...,s$ and $\mathbf{O}_{\text{train}}^i\in\mathbf{O}_{\text{train}}$. Finally, corresponding feature and output vector elements are concatenated to create the training data set as,
\begin{eqnarray}\label{eq_D_train}
D_{\text{train}}\!=\!\big\{\!\{\mathbf{F}_\text{train}^1,\!\mathbf{O}_\text{train}^1\},\{\mathbf{F}_\text{train}^2,\!\mathbf{O}_\text{train}^2\},\dots,\{\mathbf{F}_\text{train}^s,\!\mathbf{O}_\text{train}^s\}\!\big\}.
\end{eqnarray}

The objective of the training process of the proposed DNN algorithm is to optimize the weight ($\textbf{W}_k$)  and bias ($\textbf{b}_k$) parameters of the DNN so that the difference between the actual output value of the DNN and the target value is minimized. To achieve this, the weights and bias vectors of the hidden layers of the DNN are updated at each iteration during training. $\textbf{W}_k$  and $\textbf{b}_k$ are the weight and bias vectors of the $k^{th}$ layer with $k=1,2,\ldots,\mathcal{K}$, where $\mathcal{K}=5$. According to the proposed model, DNN output $\hat{\boldsymbol{\phi}}_{\ell}$ for each relay is a function of $\textbf{F}_{\text{train}}$ and $\Theta \triangleq \bigl\{\boldsymbol{\theta}_1,\boldsymbol{\theta}_2,\ldots,\boldsymbol{\theta}_{s}\bigr\}$ with $\boldsymbol{\theta}_k \triangleq \bigl\{\textbf{W}_k,\textbf{b}_k\bigr\}$, where $\Theta$ denotes the training parameter set consisting of $\textbf{W}_k$ and $\textbf{b}_k$ values. The difference between the actual output and target values is determined by the loss function which is defined for the proposed DNN architecture as, 
\begin{eqnarray}\label{eq_loss_relay}
\mathcal{L}(\Theta) = \frac{1}{2}\sum_{i=1}^s \left \| \boldsymbol{\phi}_{\ell,n}^i- \hat{\boldsymbol{\phi}}_{\ell,n}^i \big(\Theta\big) \right \|^2
\end{eqnarray}
where, $\boldsymbol{\phi}_{\ell,n}$ is the target output value, $\hat{\boldsymbol{\phi}}_{\ell,n}$ is the actual DNN output and $i$ is the index of the responses. On each iteration of the network, optimization algorithm constantly compares the loss value for each respective DNN output and optimizes $\textbf{W}_k$ and $\textbf{b}_k$ values on each layer to converge to the desired value. Optimization is carried out for each iteration based on the Adam (adaptive moment estimation optimization algorithm) optimizer, which can be defined as,
\begin{eqnarray}\label{eq_adam}
\Theta_{{\mu}+1}=\Theta_{\mu}-\frac{\eta m_{\mu}}{\sqrt{v_{\mu}+\epsilon}}
\end{eqnarray}
where $\eta$ is the learning rate determining the step size used to update the weights, $\epsilon$ is a smoothing term to avoid division by zero,  and $\mu$ is the iteration index \cite{kingma2017adam}. Adam optimizer updates the training parameter $\Theta$ at each iteration based on element-wise moving averages of both the parameter gradients and their squared values $m_{\mu}$ and $v_{\mu}$, respectively, which can be expressed as, 
\begin{eqnarray}\label{eq_adam_m}
m_{\mu}=\delta_1m_{{\mu}-1}+(1-\delta_1)\nabla\mathcal{L}(\Theta),
\end{eqnarray}
\begin{eqnarray}\label{eq_adam_v}
v_{\mu}=\delta_2m_{{\mu}-1}+(1-\delta_2)[\nabla\mathcal{L}(\Theta)]^2.
\end{eqnarray}
Here, $\delta_1$ and $\delta_2$ are the decay rates of the moving averages, and $\nabla\mathcal{L}(\Theta)$ is the gradient of the loss function given in (\ref{eq_loss_relay}). If the gradients in (\ref{eq_adam_m}) and (\ref{eq_adam_v}) stay close after several iterations, the moving averages assist for the parameter updates to gain momentum in a certain direction. On the other hand, if the gradients contain a great amount of noise, then the parameter updates will be smaller due to the moving averages of the gradients getting smaller. This mechanism of the Adam is crucial for our case since the received signals contain a great deal of channel noise, exposed to path loss and occasionally mismatched phase angles at the relay \cite{Song2021}.

\subsubsection{DNN$_{R, D}$\:-\:CRIS Model}
As the proposed DNN is supposed to be deployed at the destination, input data for the DNN will be noisy, path faded and modulated signals reaching the destination ($y_\ell$), and the output will be the estimations for original transmitted symbols ($\hat{x}$). The estimation problem for this case is a multiclass classification problem such that for every received signal $y_\ell$, the DNN based estimator should converge to the $\hat{x}$ estimate closest to the original symbol, which will be chosen from a pre-determined symbol set.

The symbol set and the number of classes to be used in the classification process are based on the modulation order, such that for $M^{th}$ modulation order, there will be $M$ symbols in the symbol set and $M$ classes consequently. In this respect, for $M^{th}$  order, the symbol set and the respective classes can be defined as,
\begin{eqnarray}\label{eq_S}
\mathcal{S}=\Big\{x_{\mathcal{S}_1},x_{\mathcal{S}_2}\ldots,x_{\mathcal{S}_M}\Big\},
\end{eqnarray}
\begin{eqnarray}\label{eq_C}
\mathcal{C}=\Big\{\mathbb{X}_1,\mathbb{X}_2,\ldots,\mathbb{X}_M\Big\}.
\end{eqnarray}
Here, $\mathcal{C}$ is basically a set of class labels assigned for each member of the symbol set $\mathcal{S}$, where $x_{\mathcal{S}_v}\in\mathcal{S}$ and $\mathbb{X}_v\in\mathcal{C}$ for $v=1,2,\ldots,M$. It can be seen that each element of $\mathcal{C}$ can be mapped to the respective element in $\mathcal{S}$ and vice versa, which ultimately results in the estimate $\hat{x}_s$ as seen in Fig. \ref{DNN_model1}. This relation between the symbol set and classes can be represented by  $\big\{x_{\mathcal{S}_v}\longleftrightarrow\mathbb{X}_v\big\}\longrightarrow{\hat{x}_s}$. In this case, feature vector and the corresponding output vector for the proposed DNN can be stated as,
\begin{eqnarray}\label{eq_F_train_dest}
\mathbf{F}_{\text{train}}^i = \Big[ \Re(\mathbf{y}_{\ell,n}), \Im(\mathbf{y}_{\ell,n})\Big]_{1\times2},
\end{eqnarray}
\begin{eqnarray}\label{eq_O_train_dest}
\mathbf{O}_{\text{train}}^i  = \Big[\mathbb{X}_v^i\Big]_{1\times1}, \quad v=1,2,\ldots,M.
\end{eqnarray}
Here, it should be noted that $\mathbf{y}_{\ell,n}$ is separated into real and imaginary parts as in (\ref{eq_F_train}) for $i=1,2,\ldots,s$ and $\mathbb{X}_v$ is a class category for $\mathbf{O}_{\text{train}}^i$. In this case, the training data set can be generated by concatenating feature vectors and corresponding symbol/class pairs for the sample space $s$ as shown in (\ref{eq_D_train}).

\begin{algorithm}[t!]
  \caption{Theoretical Training Algorithm for DNN @ $R_\ell$}
 \begin{algorithmic}[1]
 \renewcommand{\algorithmicrequire}{\textbf{Input:}}
 \renewcommand{\algorithmicensure}{\textbf{Output:}}
 \REQUIRE CSI data - $h_{\ell,n}, g_{\ell,n}$
 \ENSURE  Trained DNN at $R_\ell$ - $\textbf{W}_k$, $\textbf{b}_k$
 \\ \textit{Initialisation} :Initialize the DNN parameters - $\textbf{W}_k$, $\textbf{b}_k$, and loss $\mathcal{L}(\Theta)$ are set to zero.
 \\ \textit{LOOP Process}
  \FOR {$i = 1$ to $s$}
  \STATE Pre-process the CSI data - Real and imaginary parts of $h_{\ell,n}$ and $g_{\ell,n}$ are separated and rearranged as in (\ref{eq_F_train}), generating feature vector $\mathbf{F}_{\text{train}}^i$.
  \STATE Compute the theoretical adjusted phase vector $\boldsymbol{\phi}_{\ell,n}$ maximizing instantenous SNR as in (\ref{eq_SNR_max}).
  \STATE Pre-process the phase data found in step 3 - Real and imaginary parts of $\boldsymbol{\phi}_{\ell,n}$ are separated and rearranged as in (\ref{eq_O_train}), generating output vector $\mathbf{O}_{\text{train}}^i$.
  \ENDFOR
  \STATE Generate the training data - Feature and output vectors are concatenated to form  $D_{\text{train}}$ as in (\ref{eq_D_train}).
  \STATE Train the DNN until $\mathcal{L}(\Theta)$ is minimized with respect to (\ref{eq_loss_relay}) and (\ref{eq_adam}).
 \RETURN Trained DNN at $R_\ell$
 \end{algorithmic} 
  \end{algorithm}
	\begin{algorithm}[t!]
  \caption{Theoretical Training Algorithm for DNN @ $D$}
 \begin{algorithmic}[1]
 \renewcommand{\algorithmicrequire}{\textbf{Input:}}
 \renewcommand{\algorithmicensure}{\textbf{Output:}}
 \REQUIRE Original transmitted symbols ($x$) and noisy, path faded and modulated signals received at the destination ($\mathbf{y}_{\ell,n}$).
 \ENSURE  Trained DNN at $D$ - $\textbf{W}_k$, $\textbf{b}_k$
 \\ \textit{Initialisation} :Initialize the DNN parameters - $\textbf{W}_k$, $\textbf{b}_k$, and loss $\mathcal{L}(\Theta)$ are set to zero.
 \\ \textit{LOOP Process}
  \FOR {$i = 1$ to $s$}
  \STATE Pre-process received signals at the destination - Real and imaginary parts of $\mathbf{y}_{\ell,n}$ are separated and rearranged as in (\ref{eq_F_train_dest}), generating feature vector $\mathbf{F}_{\text{train}}^i$.
  \STATE Extract the transmitted symbols and convert symbol set into a class set as in (\ref{eq_S}) and (\ref{eq_C}), respectively.
  \STATE Generate output vectors - For every feature vector, the corresponding class is designated as an output vector $\mathbf{O}_{\text{train}}^i$ as in (\ref{eq_O_train_dest}).
  \ENDFOR
  \STATE Generate the training data - Feature and output vectors are concatenated to form  $D_{\text{train}}$ as in (\ref{eq_D_train}).
  \STATE Train the DNN until $\mathcal{L}(\Theta)$ is minimized with respect to (\ref{eq_loss_dest}) and (\ref{eq_adam}).
 \RETURN Trained DNN at $D$
 \end{algorithmic} 
  \end{algorithm}

For the multiclass classification problem, the loss function which will be applied in the process can be defined as,
\begin{eqnarray}\label{eq_loss_dest}
\mathcal{L}(\Theta) = -\frac{1}{s}\sum_{i=1}^s\sum_{v=1}^M \mathbb{X}_v^i\ln\Big({\Hat{\mathbb{X}}_v^i}\big(\Theta\big)\Big)
\end{eqnarray}
where, $s$ is the number of samples, $M$ is the number of classes, $\Hat{\mathbb{X}}_v^i$ is the actual output value for the $v^{th}$ class at $i^{th}$ sample, and $\mathbb{X}_v^i$ is the target value for the same sample. Here, $\ln(.)$ denotes the natural logarithm function. As the classification layer follows the softmax layer, $\Hat{\mathbb{X}}_v^i\big(\Theta\big)$ value will be the output of the softmax layer. (\ref{eq_loss_dest}) is also called as the cross-entropy loss between the actual and target outputs \cite{bridle-softmax}.

\begin{figure*}[t!]
\centering
\includegraphics[width=1\textwidth]{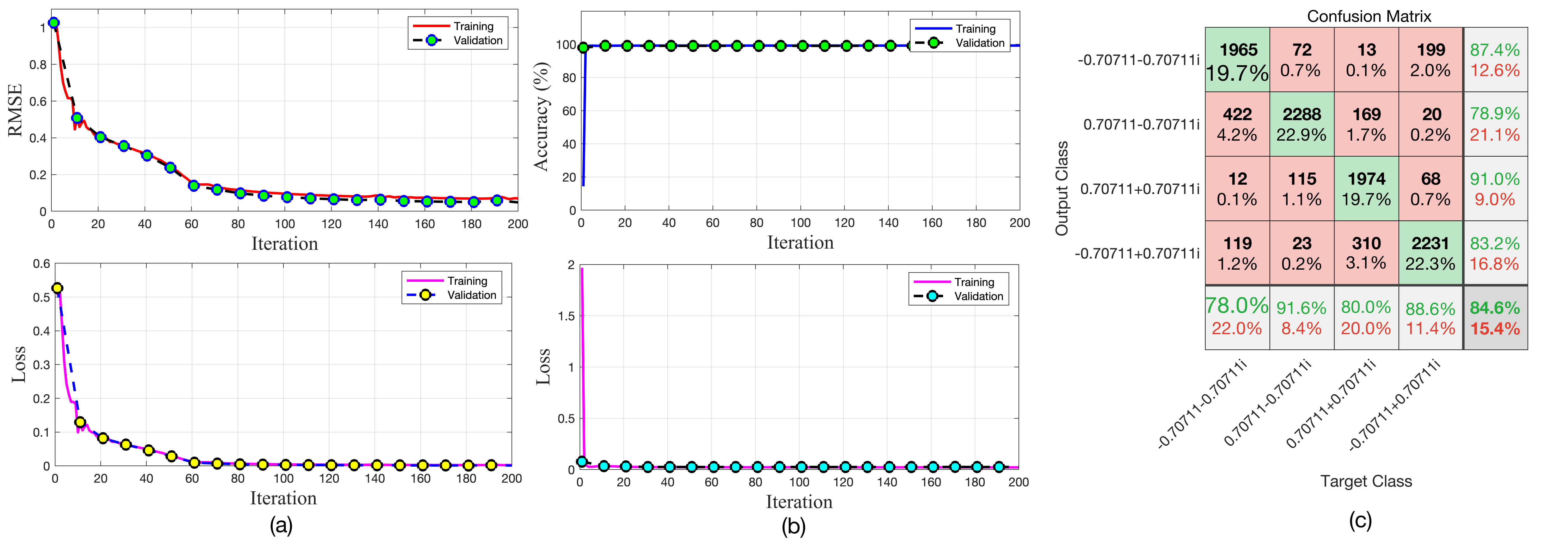}
\caption{Training Progress of the DNN (a) at $R_\ell$, in terms of RMSE and loss using ReLU, (b) at $D$, in terms of accuracy (\%) and loss using ReLU, (c) Confusion matrix for DNN at $D$ with $M=4$ for  SNR=$-40$ dB.}
\label{RMSE for relu}
\vspace{-0.3cm}
\end{figure*}

For optimizing the loss function in (\ref{eq_loss_dest}), we also deploy Adam and follow the same blueprint stated for the previous DNN as in (\ref{eq_adam}), (\ref{eq_adam_m}) and (\ref{eq_adam_v}).

Based on the training processes presented so far, theoretical algorithms for both DNNs can be shown as in Algorithm 1 and Algorithm 2.

\begin{table} [t!]
\vspace{-0.35cm}
\caption{Training Parameters.}
\begin{center}
\begin{tabular}{|c|c|c|}
 \hline
  \hline
 \textbf{Training Parameter} &  \textbf{DNN@Relay} & \textbf{DNN@Destination} \\
 \hline\hline
{Modulation Order ($M$)} &\multicolumn{2}{c|}{4/8} \\ 
 \hline
 {Num. of Transmit Antennas} &\multicolumn{2}{c|}{1} \\
 \hline
 {Num. of Reflect. Elements ($N$)} &\multicolumn{2}{c|}{8/32} \\
 \hline
 {Pathloss Exponent ($c$)} &\multicolumn{2}{c|}{4} \\
 \hline
 {Num. of Hidden Layers} &\multicolumn{2}{c|}{4} \\
 \hline
{Batch Size} &\multicolumn{2}{c|}{256} \\
 \hline
  {Validation Split} &\multicolumn{2}{c|}{10\%} \\ 
 \hline
 {Num. of Samples ($s$)} & 360000 & 45000 \\ 
 \hline
 {Iteration Steps} & 300 & 150 \\ 
 \hline
{Learning Rate ($\eta$)} & 0.003 & 0.003 \\ 

 \hline
  \hline
\end{tabular}
\label{table:2}
\end{center}
\vspace{-0.4cm}
\end{table}

\vspace{-0.4cm}
\section{Numerical Results and Discussions}


\subsection{Simulation Parameters for DNN$_R$\:-\:CRIS/DNN$_{R, D}$\:-\:CRIS}

Simulation is based on the channel and system model described in Section II, where $S$, $R_{\ell}$, and $D$ have been positioned as depicted in Fig. \ref{S_model1}. All relays are kept within the half circle area by choosing $ \pi/2<\theta_\ell<\pi$, with $d_{SD}$, $d_{SR_{\ell}}$, and $d_{R_{\ell}D}$ distances normalized with respect to $d_{SD}=1$. The SNR used in the simulations herein is defined as $\mathrm{SNR (dB)}=10\log_{10}(E_s/N_0)$. The number of reflecting elements on RIS based relays is originally specified between 8 and 128 for observing its effect on relay configurations and BER performances, however, due to the space constraints, plots for 64 and 128 reflectors haven't been included in the manuscript as the results do not show a remarkable deviation from overall trend. All channels are subject to Rayleigh fading and path loss effects with path loss exponent $c$ chosen as $4$ for Fig. \ref{BER R2N8} - Fig. \ref{BER R2N8_v2}, and $c=3$ for Fig. \ref{distances} \cite{Rappaport1992}.

We use MATLAB and its toolboxes for the simulation with the basic training parameters given in Table \ref{table:2}. As stated in Section III.A, ReLU activation function is opted for deployment in hidden layers for both DNN models in all simulation phases. During training, $10\%$ of the training data set is reserved for validation by the validation split and validation frequency is set as $10$ for both DNNs. Another critical parameter for the training phase is the  batch size, which determines the size of the data batch -also called as a mini-batch- processed by the network on each iteration. Batch size is basically a subset of the training data set, and for both DNN models, it is set as 256.


\subsection{DNN Performance Analysis}
Performances of the DNNs deployed in the proposed model have been evaluated using a number of tools during training and transmission phases.

Fig. \ref{RMSE for relu}a and \ref{RMSE for relu}b show the training progress plots for DNNs at $R$ and $D$, respectively, providing information on the training and validation performance in terms of root mean square error (RMSE) and loss values for DNN at $R$, and accuracy percentage for DNN at $D$. These metrics are calculated during training for each mini-batch, processed by the network on each iteration. As it can be observed from the training and validation curves, both DNNs showed a good generalization performance with little to no need for extra measures for overfitting.



Loss values for the proposed DNNs at $R$ and $D$ can be directly computed by the functions defined in (\ref{eq_loss_relay}) and (\ref{eq_loss_dest}), respectively. On the other hand, RMSE for a typical regression network -DNN at $R_\ell$ in our case, can be defined as,
\begin{eqnarray}\label{eq_RMSE}
RMSE=\sqrt{\frac{1}{s}\sum_{i=1}^s(t_i-\hat{t}_i)^2}.
\end{eqnarray}
Here, $t_i$ is the target output, $\hat{t}_i$ is the actual output, and $s$ denotes the number of samples or responses in general terms \cite{Kay97}. RMSE and loss curves in a regression network will follow a similar path as loss function is half mean squared error computed by the network for optimizing the parameters.

The accuracy of the DNN at $D$ can also be monitored during the transmission phase, using confusion matrices. The confusion matrix for classification networks shows the percentage of the actual DNN outputs matching with the target outputs for each prediction cycle of the network. Fig. \ref{RMSE for relu}c shows the confusion matrix for DNN deployed at $D$ for $M=4$ and SNR=$-40$ dB. Here, the horizontal and vertical axes show the target output and actual output, respectively. The cells on the green diagonal show the percentage of the successfully matched symbols for each symbol class, and the cell at the lower right corner shows the overall percentage  of accuracy for all symbol classes. During simulation, the DNN output accuracy has a direct impact on the BER performance, such that a high estimation accuracy for symbol detection improves BER values considerably.

During the simulation, both DNNs at the relay and destination have been trained for a range of SNR levels for optimum accuracy, and we observed the generalization ability of the algorithm when the training and transmission SNRs were different. It is found that the general accuracy level improves with the increase in SNR as expected and has similar results for different training and transmission SNRs.

\vspace{-0.3cm}

\subsection{Complexity Analysis}

In this section, we will analyze and compare the computational complexities of RS, MRC, and ML-based receivers against the DNN-assisted symbol detection algorithm deployed in DNN$_{R, D}$\:-\:CRIS model.

The complexity analysis for the proposed DNN model is based on the number of operations in each layer. As the input neurons function as the entry point for the signals and do not perform any operation, they will not be considered in the analysis as a layer.

\begin{table}[t!]
\centering
\caption{Computational Complexity Analysis.}
\label{table:3}
\begin{tabular}{| c | c | c | P{1.8cm} |}
 \hline
 \hline
 \textbf{RS} & \textbf{MRC} & \textbf{ML} & \textbf{DNN} \\
 \hline\hline
 \!\!\!\!$\mathcal{O}\Big(N+4M\Big)$ \!\!\!\!&\!\!\!\!\! $\mathcal{O}\Big(L(N+1)+4M\Big)$ \!\!\!\!&\!\!\!\!\! $\mathcal{O}\Big(LN+4L^2M\Big)$\!\! &\!\! 
$\mathcal{O}\Big(pn_1+n_{\mathcal{K}}o +\sum_{k=1}^{\mathcal{K}-1}n_kn_{k+1}\Big)$
 \\
 \hline
 \hline
\end{tabular}
\end{table}

\begin{table}
\caption{Parameters of Complexity Scenarios and DNN Configurations.}
\begin{center}
\label{table:4}
\begin{tabular}{|p{3.2cm}|c|c|c|}
 \hline
  \hline
 \textbf{} &  \textbf{Scenario 1} & \textbf{Scenario 2} & \textbf{Scenario 3}\\
 \hline\hline
 {Num. of Relays ($L$)} & 4 & 6 & 24 \\ 
 \hline
{Modulation Order ($M$)} & 4 & 8 & 16 \\ 
 \hline
 {Num. of Refl. Elements ($N$)} & 16 & 32 & 128  \\
 \hline
 {DNN-1} &\multicolumn{3}{c|}{$\mathcal{K}=4$, $n_1=n_2=n_3=n_4=256$}\\ 
 \hline
 {DNN-2} &\multicolumn{3}{c|}{$\mathcal{K}=2$, $n_1=n_2=16$}\\ 
 \hline
{DNN-3} &\multicolumn{3}{c|}{$\mathcal{K}=2$, $n_1=n_2=8$}\\ 

 \hline
  \hline
\end{tabular}
\end{center}
\vspace{-0.3cm}
\end{table}

In each neuron, a mapping function is performed as in (\ref{eq_f_k}), which involves the linear combination of the input with neuron parameters $\textbf{W}_k$, $\textbf{b}_k$ and later processed by the activation function. In a multi-layer neural network, the complexity depends on the number of layers and the size of each layer, both of which scale the number of vector multiplications. Hence, for a multi-layer neural network with $\mathcal{K}$ hidden layers and $p$ inputs, if the $k^{th}$ hidden layer has $n_k$ neurons with $o$ neurons at the output, the total number of multiplications will be,
\begin{eqnarray}\label{eq_complexity_DNN}
\mathcal{O}\Big(pn_1+on_{\mathcal{K}} +\sum_{k=1}^{\mathcal{K}-1}n_kn_{k+1}\Big),
\end{eqnarray}
where all the layers are assumed to be fully connected.

\begin{figure}[t!]
\centering
\includegraphics[width=0.48\textwidth]{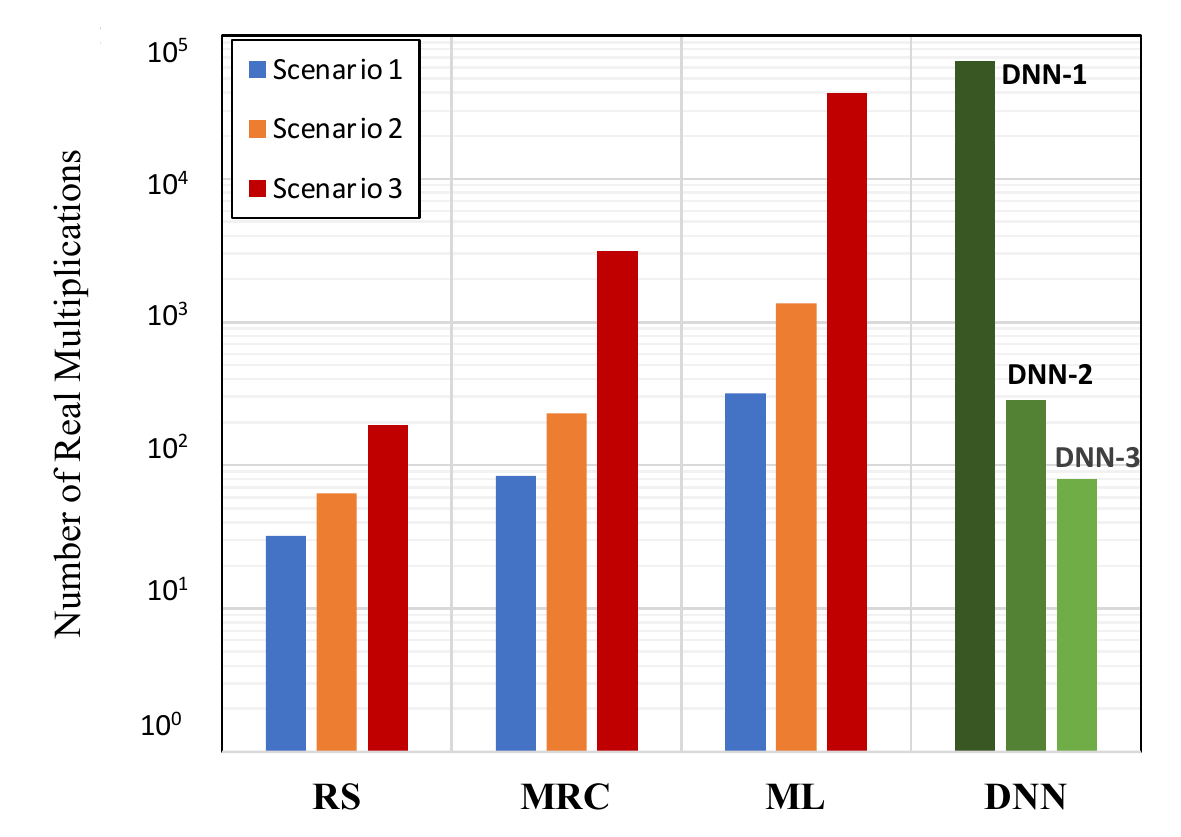}
\caption{Complexity comparisons of RS, MRC, ML, and DNN-based CRIS systems.}
\label{complexity_graph}
\vspace{-0.3cm}
\end{figure}

The computational complexities of RS, MRC, ML, and DNN-based systems in terms of number of real multiplications are illustrated in Table \ref{table:3}. For RS, MRC, and ML-based detectors, we evaluated three scenarios with various $L$, $M$, and $N$ values as given in Table \ref{table:4} and compared them against the DNN-based detectors with different layer and neuron configurations. It can be observed from the Fig. \ref{complexity_graph} that the complexities of RS, MRC, and ML-based detectors increase as the $L$, $M$, and $N$ usage grow in a large scale network. In this regard, ML and RS are found to be the most and least computationally complex detection schemes, respectively, where the relay usage is the most decisive element for both. On the DNN side, the complexity depends solely on the layer and neuron configurations. Various DNN configurations are evaluated for comparison as given in Table \ref{table:4}, all of which exhibit nearly identical detection accuracy during simulations. It is found that the complexity can be reduced successfully to a minimum with DNN-3 configuration, where the complexity is lower than all ML and MRC scenarios and even lower than RS scenarios with a high number of network components.

\begin{figure}[t!]
\centering
\includegraphics[width=0.48\textwidth]{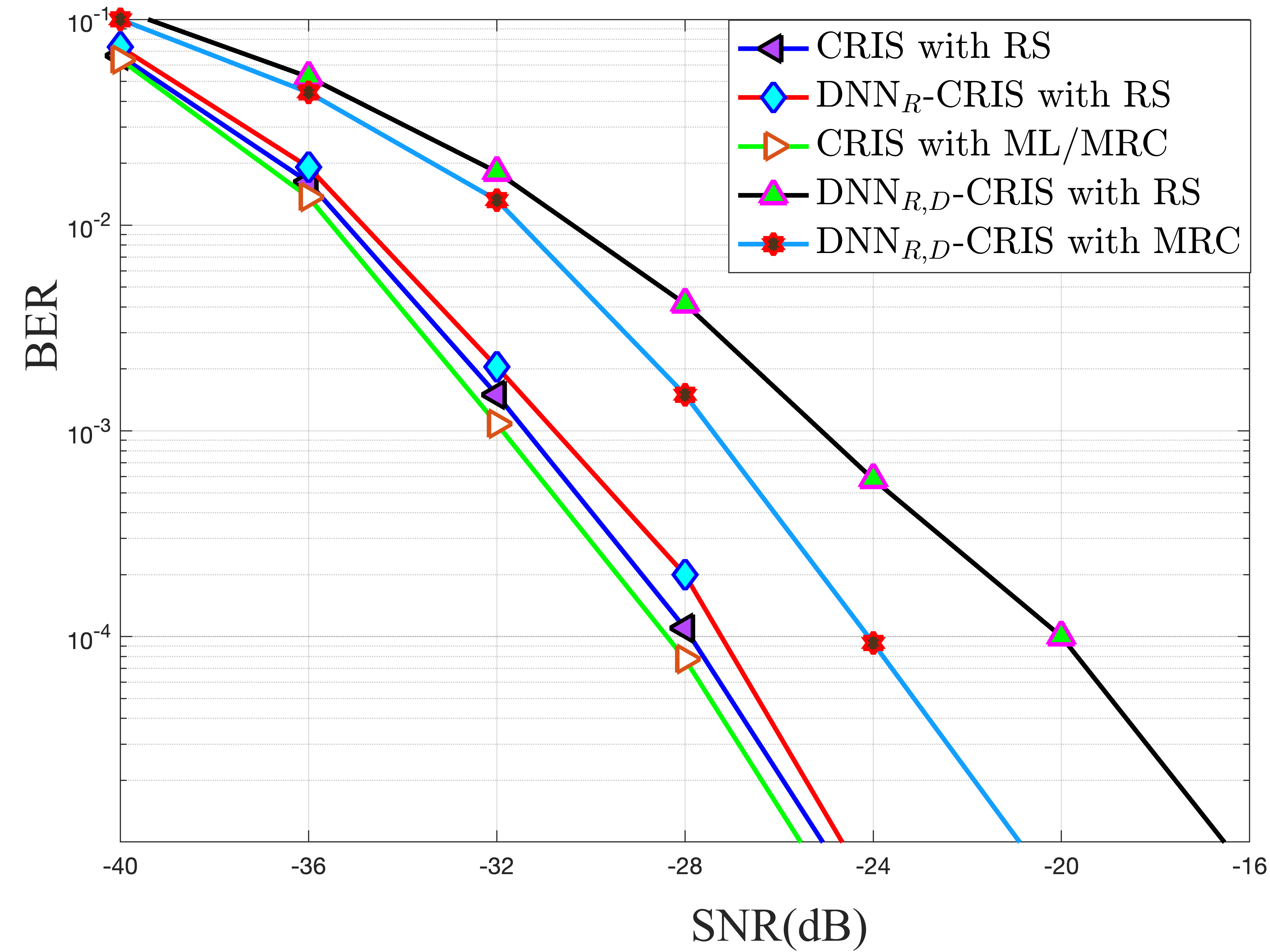}
\caption{BER comparisons of DNN$_R$-CRIS with RS, DNN$_{R, D}$-CRIS with RS/MRC, and CRIS with RS/ML/MRC systems for $M=4, L=2$ and $N_1=N_2=8$.}
\label{BER R2N8}
\end{figure}



\subsection{BER Analysis of DNN$_R$\:-\:CRIS/DNN$_{R, D}$\:-\:CRIS Schemes}
In this section, we will evaluate the BER performances of the DNN$_R$\:-\:CRIS and DNN$_{R, D}$\:-\:CRIS models using various relay and receiver configurations, including  non-DNN RIS-based relay configurations (CRIS) with RS, MRC and ML-based receivers at $D$ as a benchmark for all BER performance comparisons. For all the models with RS scheme, only the best performing relay is depicted in the figures.


\begin{figure}[t!]
\centering
\includegraphics[width=0.48\textwidth]{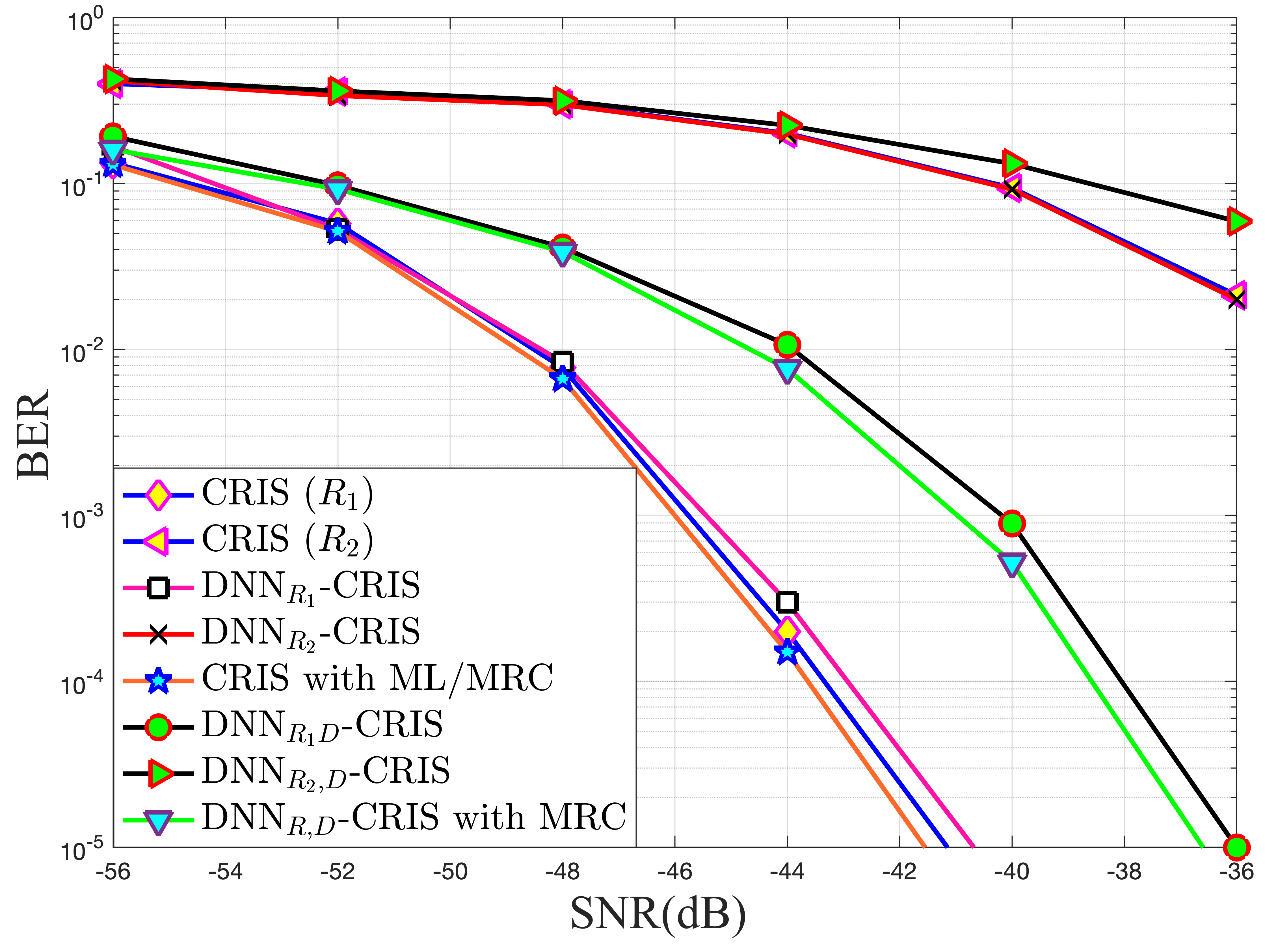}
\caption{BER comparisons of DNN$_R$-CRIS/DNN$_{R, D}$-CRIS and CRIS systems for $M=4, L=2$ and $N_1=N_2=32$.}
\label{BER R2N32}
\end{figure}

We start evaluating the BER performance of a two-relay cooperative system, using a 4-QAM modulation scheme with RS. Each RIS-based relay contains 8 reflecting elements with relay distances $d_{SR_{1}}=0.2$, $d_{R_1D}=0.98$ and $d_{SR_2}=0.5$, $d_{R_2D}=0.86$  for $\theta_1=\theta_2=\pi/2$. In this configuration, $R_1$ is selected as the best performing relay according to (\ref{eq_BER_rs}), which is also the closest relay to $S$, and its BER performance is depicted in Fig. \ref{BER R2N8}. It can be seen that performance of the DNN$_R$\:-\:CRIS model is very close to CRIS. The addition of DNN-assisted symbol estimation into the system (DNN$_{R, D}$\:-\:CRIS) introduces a slight impact on the BER performance as expected, such that, for a BER level of $10^{-3}$ the SNR loss is about $6$ dB. It is also noteworthy that the DNN$_{R, D}$\:-\:CRIS with MRC shows comparable performance with CRIS with MRC system, even outperforming non-DNN reference configurations.

In Fig. \ref{BER R2N32}, the number of reflecting elements is increased to 32, with all remaining parameters retained from the previous scenario. The immediate effect of increasing reflecting elements is observed as an SNR improvement of about $7$ dB for $N=32$ at the BER of $10^{-3}$ or below. We see that configurations with the DNN$_R$\:-\:CRIS model show nearly identical performance with non-DNN reference configurations, and the impact of DNN-based destination on SNR remains at similar levels. Increasing the number of reflecting elements on the relays improves the BER performances of DNN-based systems as a whole, preserving their performance characteristics nevertheless.

\begin{figure}[t!]
\centering
\includegraphics[width=0.48\textwidth]{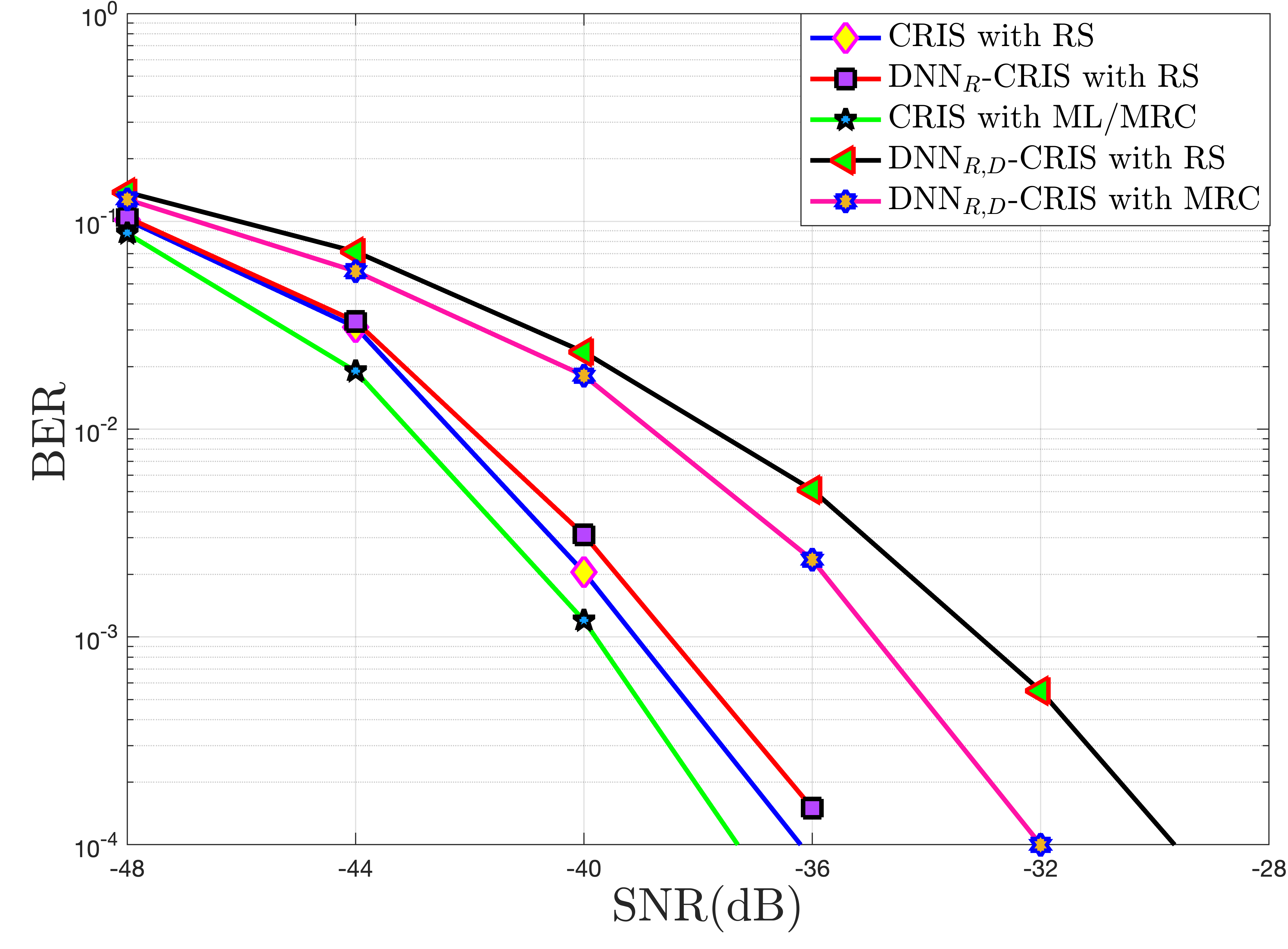}
\caption{BER comparisons of DNN$_R$-CRIS with RS, DNN$_{R, D}$-CRIS with RS/MRC, and CRIS with RS/ML/MRC systems  for $M=4, L=4$ and $N_1\!=\!N_2\!=\!N_3\!=\!N_4\!=\!8$.}
\label{BER R4N64}
\end{figure}

\begin{figure}[t!]
\centering
\includegraphics[width=0.48\textwidth]{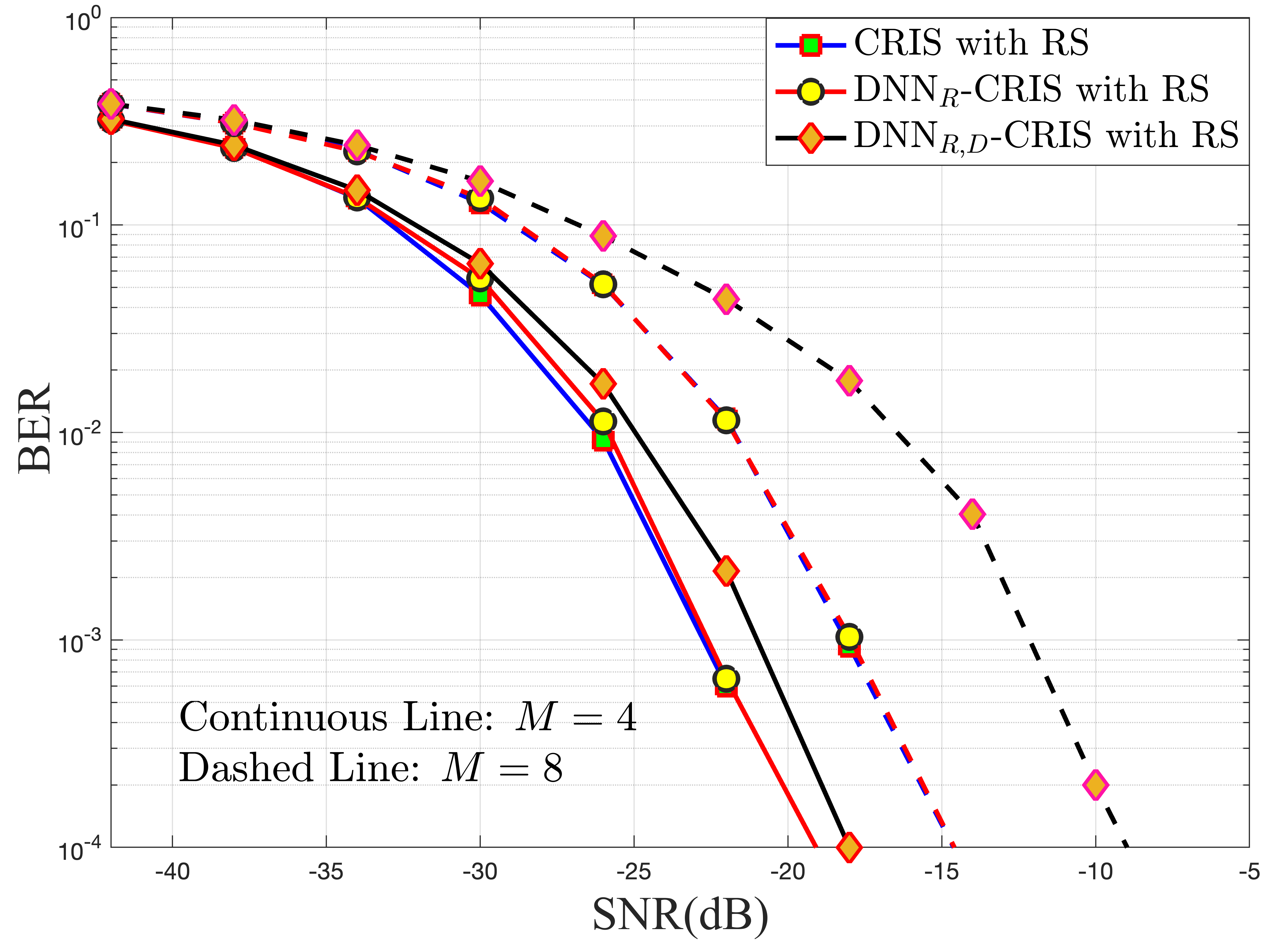}
\caption{BER comparisons of DNN$_R$-CRIS with RS, DNN$_{R, D}$-CRIS with RS, and CRIS with RS systems  for $M=4,8, L=2$ and $N_1=N_2=8$.}
\label{BER R2N8_v2}
	\vspace{-0.3cm}
\end{figure}


Fig. \ref{BER R4N64} shows an RS outcome among a four-relay system with relay distances of $d_{SR_{1}}=0.4$, $d_{R_1D}=0.91$, $d_{SR_2}=0.55$, $d_{R_2D}=0.83$,  $d_{SR_{3}}=0.65$, $d_{R_3D}=0.76$,  $d_{SR_4}=0.2$, and $d_{R_4D}=0.98$, each $R$ having 8 reflecting elements. Here, as in Fig. \ref{BER R2N8}, the best-performing relay is the closest one to $S$, which is $R_4$ in this case. Once more, we observe that the DNN$_R$\:-\:CRIS scenario shows nearly identical BER performance with non-DNN configuration as in the previous cases. Here, the path loss effects become clearer as the relays are distributed throughout the model layout, such that configurations incorporating relays closer to $S$ or $D$ show a better BER performance as the signals are exposed to a lesser attenuation. The most significant outcome here is that the BER value is found to be maximum when $d_{SR_\ell}=d_{R_{\ell}D}$.

We also compared the BER performances for $M=4$ and $M=8$ schemes as shown in Fig. \ref{BER R2N8_v2}, using RS among the relays with distances $d_{SR_{1}}=0.9$, $d_{R_1D}=0.43$, $d_{SR_{2}}=0.35$ and $d_{R_2D}=0.93$. It can be seen that, for both modulation schemes, the performance characteristics remain mostly unchanged as the best performance is achieved for $R_1$, the closest one to $D$, where the DNN-based destination introduced a certain amount of SNR loss. The only remarkable outcome is the general performance setback for $M=8$ as expected, such that at the BER of $10^{-3}$, the SNR loss is about $5$ dB for DNN$_R$\:-\:CRIS and $8$ dB for DNN$_{R, D}$\:-\:CRIS models.

\begin{figure}[t!]
\centering
\includegraphics[width=0.48\textwidth]{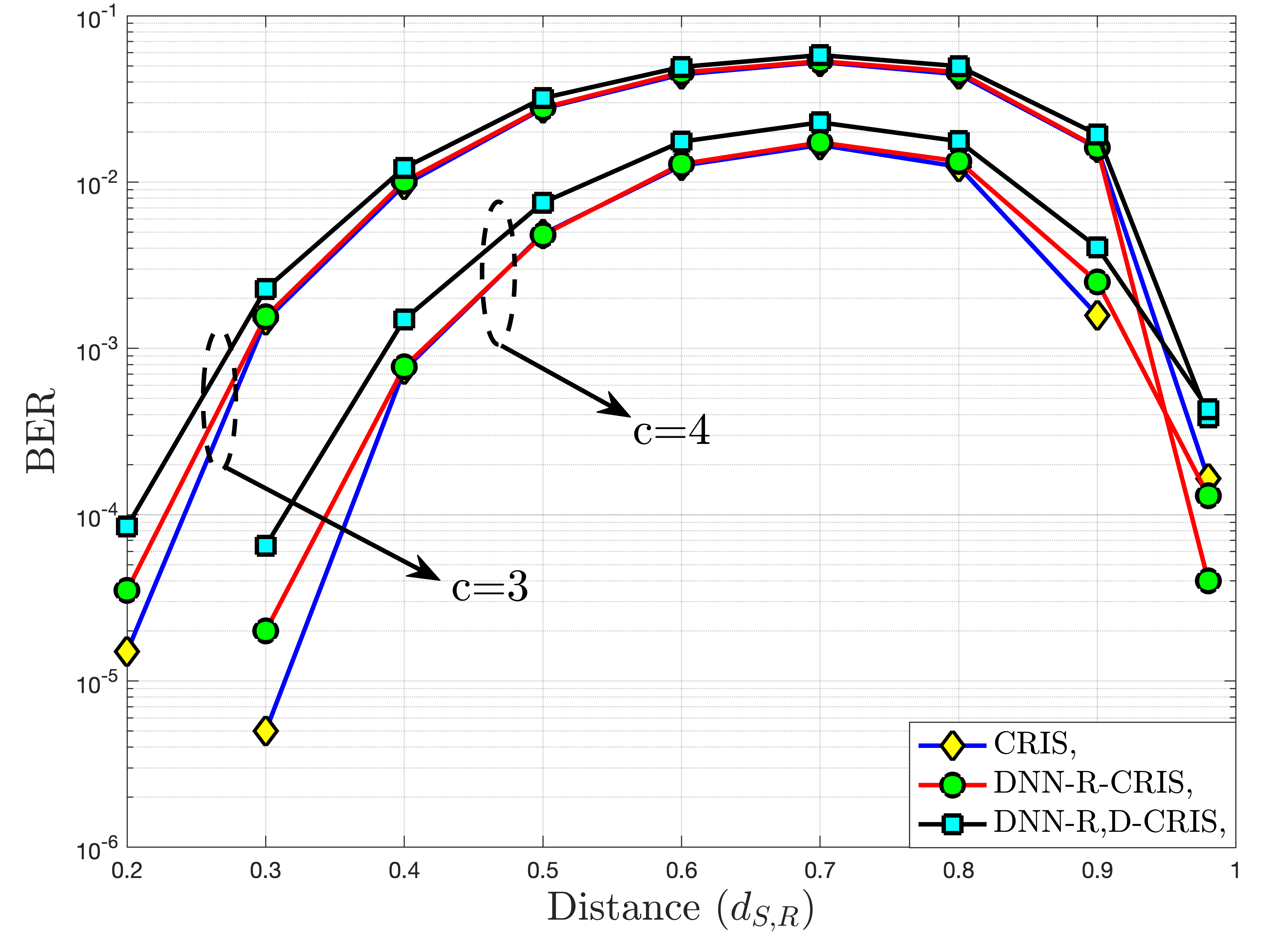}
\caption{BER vs. $d_{SR}$ comparisons of DNN$_R$-CRIS/DNN$_{R, D}$-CRIS and CRIS systems with a single relay for $M=4, N=8$, SNR=-20 when $c=3,4$.}
\label{distances}
\end{figure}

Finally, in Fig.\ref{distances}, we can observe the effect of distance on the BER performance clearly, as the worst performance is shown at $d_{SR}=0.7$ for all configurations, which is close to midpoint between $S$ and $R$. Furthermore, the effect of path loss exponent ($c$) can also be observed on the BER performance, such that increasing $c$ results in a better BER performance. This outcome is due to the fact that all the distance values are normalized with respect to $d_{SD}=1$, where $0<d_{SR_\ell},d_{R_{\ell}}\leq1$.

\section{Conclusion and Future Works}
In this work, we realize a deep learning-assisted RIS-based cooperative communication system with a DNN-based symbol detector at the destination, where all the RIS elements are considered to be interconnected through an IoT network. The effect of the path loss on the DNN$_R$\:-\:CRIS and DNN$_{R, D}$\:-\:CRIS models is demonstrated in terms of BER performance. We incorporate DNN performance analysis techniques during the training and/or transmission phases for evaluating the individual DNN performance and analyze the BER performance to evaluate the total system performance in all deployment scenarios. The simulation results show that DNN-assisted RIS-based relays can effectively maximize the reflected signals and exhibit satisfactory performance against the reference non-DNN configurations in general. Combining DNN-assisted relays with DNN-based symbol detectors at the destination also obtains acceptable performance with low system complexity, especially in conjunction with RIS-based relays that harbor large numbers of reflectors. Moreover, using MRC at the destination to combine multi-relay signals for the dual DNN scenarios yields promising results, and applying several relay selection schemes to the final model is also possible.

In the context of this work, it is assumed that the CSI is perfectly known at the relay. It will be a broad and challenging field of study to consider similar scenarios under totally blind or imperfect CSI conditions. The performance of similar systems with imperfect CSI are already discussed in \cite{Basar3}, \cite{Khan2020DeepLearningaidedDF}, \cite{EfficientMIMOdetection_chen} and we expect to extend some of the methods proposed in these works in our future studies where more advanced deep learning techniques and DNN architectures will presumably be needed.


%





\ifCLASSOPTIONcaptionsoff
  \newpage
\fi


%

\bibliographystyle{ieeetr}
\bibliography{Referanslar}



%








\end{document}